\documentclass[onecolumn, showpacs,
 floatfix,aps,nofootinbib]{revtex4}
\usepackage{amssymb,amsmath}
\usepackage[dvips]{graphics,color}
\usepackage{epsfig}\usepackage{float}

\begin{document}

\title{ELKO, flagpole and flag-dipole spinor fields, and the instanton Hopf fibration}

\author{R. da Rocha}
\email{roldao.rocha@ufabc.edu.br} \affiliation{ Centro de
Matem\'atica, Computa\c c\~ao e Cogni\c c\~ao, Universidade
Federal do ABC, 09210-170, Santo Andr\'e, SP, Brazil}
\author{J. M. Hoff da Silva}
\email{hoff@ift.unesp.br} \affiliation{Instituto de F\'{\i}sica
Te\'orica, Universidade Estadual Paulista, Rua Pamplona 145
01405-900 S\~ao Paulo, SP, Brazil}


\begin{abstract}
In a previous paper we explicitly constructed a mapping that leads Dirac
spinor fields to the dual-helicity eigenspinors of the charge conjugation
operator (ELKO spinor fields). ELKO spinor fields are prime candidates for
describing dark matter, and belong to a wider class of spinor fields, the
so-called flagpole spinor fields, corresponding to the class-(5), according
to Lounesto spinor field classification, based on the relations and values
taken by their associated bilinear covariants. Such a mapping between Dirac
and ELKO spinor fields was obtained in an attempt to extend the Standard
Model in order to encompass dark matter. Now we prove that such a mapping,
analogous to the instanton Hopf fibration map $S^3\ldots S^7\rightarrow S^4$, 
prevents ELKO to describe the instanton, giving a suitable physical interpretation 
to ELKO. We review
ELKO spinor fields as type-(5) spinor fields under the Lounesto spinor field
classification, explicitly computing the associated bilinear covariants.
This paper is also devoted to investigate some formal aspects of the
flag-dipole spinor fields, which correspond to the class-(4) under the
Lounesto spinor field classification. In addition, we prove that
type-(4) spinor fields --- corresponding to flag-dipoles --- and ELKO spinor
fields --- corresponding to flagpoles --- can also be entirely described in
terms of the Majorana and Weyl spinor fields. After all, by choosing a
projection endomorphism of the spacetime algebra $\mathcal{C}\ell_{1,3}$ it
is shown how to obtain ELKO, flagpole, Majorana and Weyl spinor fields,
respectively corresponding to type-(5) and -(6) spinor fields, uniquely from
limiting cases of a type-(4) --- flag-dipole --- spinor field, in a similar
result obtained by Lounesto.
\end{abstract}

\maketitle
\tableofcontents





\section{Introduction}

ELKO --- \emph{Eigenspinoren des Ladungskonjugationsoperators} --- spinor
fields\footnote{%
ELKO is the German acronym for Dual-helicity eigenspinors of the charge
conjugation operator \cite{allu}.} represent an extended set of Majorana
spinor fields, describing a non-standard Wigner class of fermions, in which
the charge conjugation and the parity operators commute, rather than
anticommute \cite{allu,alu2,alu4,ahlu4}. Although in the algebraic framework
there is no essential difference between ELKO and Majorana spinor fields (in
the Lounesto spinor field classification), from the physical point of view
Ahluwalia and Grumiller showed in \cite{allu} that ELKO spinor fields are
competing candidates for the Majorana fields. ELKO spinor fields carry mass
dimension one, and not three-halves, and consequently cannot be part of the
SU(2)$_L$ doublets of the Standard Model, which includes spin-1/2 particles
of mass dimension three-halves. Besides, a quantum field theory constructed
for ELKO spinor fields gives a non-local character to ELKO. Ahluwalia and
Grumiller argued that what localizes otherwise extended field configurations
like solitons is a conserved topological charge, and in the absence of it,
there is nothing that protects the particle from spreading \cite{allu}.
Non-locality is related to a classical field (soliton) configuration, but
the non-locality which the current papers about ELKO refer, appears at the
level of the field anticommutators. Concerning the fundamental
anticommutators for the ELKO quantum field, such a non-locality is at the
second order, in the sense that while the field--momentum anticommutator
exhibits the usual form expected of a local quantum field theory, the
field--field and momentum--momentum anticommutators do not vanish \cite{allu}%
. In addition, the vacuum expectation value is computed to be non-trivial 
\cite{allu}. For more details, see also \cite{alu2,alu4,ahlu4}.

Further, ELKO 
accomplishes dual-helicity eigenspinors of the spin-1/2 charge conjugation
operator, and carry mass dimension one, besides having non-local properties 
\cite{allu,alu2,alu4,ahlu4}. 

It is well known that all \textit{spinors} in Minkowski spacetime can be
given --- from the classical viewpoint\footnote{%
It is well known that spinors have three different, although equivalent,
definitions: the operator, the classical and the algebraic one \cite%
{benn,moro,rod,cru,chev,op,hes,hes1,cartan,riesz}.} --- as elements of the
carrier spaces of the $D^{(1/2,0)}\oplus D^{(0,1/2)}$ or $D^{(1/2,0)}$, or $%
D^{(0,1/2)}$ representations of SL(2,$\mathbb{C)}$. \ However, according to 
\cite{allu,alu2} only in the low-energy limit, \textit{ELKO spinor fields}
carries a representation of the Lorentz group. P. Lounesto, in the
classification of spinor fields, proved that any spinor field belongs to one
of the six (disjoint) classes found by him \cite{lou1,lou2}. Such an \textit{%
algebraic} classification is based on the values assumed by their bilinear
covariants, the Fierz identities, aggregates and boomerangs \cite%
{lou1,lou2,meu}. Lounesto spinor field classification has wide applications
in cosmology and astrophysics (via ELKO, for instance see \cite%
{allu,alu2,meu,alu3,boe1}), and in General Relativity: it was recently
demonstrated that Einstein-Hilbert, the Einstein-Palatini, and the Holst%
\footnote{%
The Holst action is shown to be equivalent to the Ashtekar formulation of
Quantum Gravity \cite{hor}.} actions can be derived from the quadratic
spinor lagrangian (that describes supergravity) \cite{tn1,bars1}, when the
three classes of Dirac spinor fields, under Lounesto spinor field
classification, are considered \cite{jpe,ro3,ro4,ro5}. It was also shown 
\cite{meu} that ELKO represents the class-(5), which also incorporates
Majorana spinor fields, and that those spinor fields covers one of the six
disjoint classes in Lounesto spinor field classification. Although in \cite%
{meu} it was not found any \emph{algebraic} difference between ELKO and
Majorana spinor fields, physically ELKO describes spin-1/2 fields presenting
mass dimension one.

Any invertible map that takes Dirac spinor fields and leads to ELKO is also
capable to make mass dimension transmutations, since Dirac spinor fields
present mass dimension three-halves, instead of the mass dimension one
associated with ELKO. The main physical motivation of a previous paper \cite%
{osmano} \footnote{%
R. da Rocha thanks to Prof. Dharamvir Ahluwalia-Khalilova for private
communication on the subject.} was to provide the initial pre-requisites to
construct a natural extension of the Standard Model (SM) in order to
incorporate ELKO, and consequently a possible description of dark matter 
\cite{allu,alu2,alu3} in this context. The explicit application describing
the mapping between ELKO and Dirac spinor fields obtained in \cite{osmano}
presents an analogy to the instanton Hopf fibration map $S^3\ldots S^7
\rightarrow S^4$ mapping obtained in \cite{jayme,jayme1,bala}, and could be
interpreted as the geometric meaning of the mass dimension-transmuting
operator obtained in \cite{osmano}. It would suggest the reason why the ELKO
spinor fields satisfy the Klein-Gordon equation, instead of the Dirac
equation, as detailed and extensively shown in \cite{allu}. Physically, as
ELKO presents mass dimension one \cite{allu,alu2,alu3}, while any other type
of spin-1/2 spinor field presents mass dimension 3/2, the conditions
obtained in Sec. \ref{elkoo} might introduce the geometric explanation for
this physical open problem.

In this paper, we are also interested in geometric and algebraic properties
of flag-dipole (type-(4)) spinor fields. It is shown that such a spinor
field can be written as a ``sum'' of Weyl and Majorana spinor fields%
\footnote{%
Remember that the class of the spinor, in the Lounesto's sense, is not
necessarily preserved upon sum.}. In fact, such systematization concerning
type-(4) spinor fields can be relevant in physics, since it can be related
to the quark confinement problem \cite{lou1}. It is also shown how to obtain
type-(5) and -(6) spinor fields as limiting cases of type-(4) spinor fields.
Such results are based on some previous results already obtained by Lounesto 
\cite{lou1,lou2}, here used also to extend it to the ELKO algebraic and
geometric formalism.

The paper is organized as follows: after briefly presenting some essential
algebraic and geometric preliminaries in Section II, we introduce in Section
III the bilinear covariants together with the Fierz identities. Also, the
Lounesto classification of spinor fields is presented together with the
definition of ELKO spinor fields \cite{allu}, showing that ELKO is indeed a
flagpole spinor field with opposite (dual) helicities \cite{allu,alu2,meu}.
We carefully show the computations leading to the classification of ELKO
spinor fields as flagpole spinor fields, in the class-(5) under the Lounesto
spinor field classification, for the first time proved in \cite{meu}. In
Section IV the spacetime algebra and the construction of ideal and operator
spinor fields is reviewed, in order to the introduce the investigation of
the geometric and algebraic aspects related to the flag-dipole spinor fields
in Section VI. In Section V the instanton Hopf fibration $S^3\ldots
S^7\rightarrow S^4$ is discussed in the light of the bilinear covariants
related to Dirac spinor fields, and the relationship between such a map and
the mapping that leads Dirac spinor fields to ELKO spinor fields is
considered to show that the instanton cannot be described by an ELKO. Some
physical consequences are also discussed in this context. In the Appendix A
the definition of operator spinors is recapitulated.

\section{Preliminaries}

\label{w2}

Let $V$ be a finite $n$-dimensional real vector space and $V^{\ast }$
denotes its dual. We consider the tensor algebra $\oplus _{i=0}^{\infty
}T^{i}(V)$ from which we restrict our attention to the space $\Lambda
(V)=\oplus _{k=0}^{n}\Lambda ^{k}(V)$ of \textit{multivectors\footnote{$%
\Lambda (V^{\ast })=\oplus _{k=0}^{n}\Lambda ^{k}(V^{\ast })$ denotes the
space of the antisymmetric $k$-cotensors, isomorphic to the $k$-forms vector
space.}} over $V$. $\Lambda ^{k}(V)$ denotes the space of the antisymmetric $%
k$-tensors. Given $\psi \in \Lambda (V)$, $\tilde{\psi}$ denotes the \emph{%
reversion}, an algebra antiautomorphism given by $\tilde{\psi}%
=(-1)^{[k/2]}\psi $ ([$k$] denotes the integer part of $k$). If $V$ is
endowed with a non-degenerate, symmetric, bilinear map ${\mbox{\bfseries\slshape g}}:V\times
V\rightarrow \mathbb{R}$, it is possible to extend ${\mbox{\bfseries\slshape g}}$ to $\Lambda
(V)$. Given $\psi ={\mathbf{u}}^{1}\wedge \cdots \wedge {\mathbf{u}}^{k}$
and $\phi ={\mathbf{v}}^{1}\wedge \cdots \wedge {\mathbf{v}}^{l}$, for ${%
\mathbf{u}}^{i},{\mathbf{v}}^{j}\in V$, one defines $\mbox{\bfseries\slshape g}(\psi ,\phi )=\det (%
{\mbox{\bfseries\slshape g}}({\mathbf{u}}^{i},{\mathbf{v}}^{j}))$ if $k=l$ and ${\mbox{\bfseries\slshape g}}%
(\psi ,\phi )=0$ if $k\neq l$. The projection of a multivector $\psi =\psi
_{0}+\psi _{1}+\cdots +\psi _{n}$, $\psi _{k}\in \Lambda ^{k}(V)$, on its $p$%
-vector part is given by $\langle \psi \rangle _{p}$ = $\psi _{p}$. Given $%
\psi ,\phi ,\xi \in \Lambda (V)$, the \textit{left contraction} is defined
implicitly by ${\mbox{\bfseries\slshape g}}(\psi \lrcorner \phi ,\xi )={\mbox{\bfseries\slshape g}}(\phi ,\tilde{%
\psi}\wedge \xi )$. For $a\in \mathbb{R}$, it follows that $\mathbf{v}%
\lrcorner a=0$. The \textit{right contraction} is analogously defined by $%
{\mbox{\bfseries\slshape g}}(\psi \llcorner \phi ,\xi )={\mbox{\bfseries\slshape g}}(\phi ,\psi \wedge \tilde{\xi%
})$. Both contractions are related by $\mathbf{v}\lrcorner \psi =-\hat{\psi}%
\llcorner \mathbf{v}$. The Clifford product between ${\mathbf{w}}\in V$ and $%
\psi \in \Lambda (V)$ is given by ${\mathbf{w}}\psi ={\mathbf{w}}\wedge \psi
+{\mathbf{w}}\lrcorner \psi $. The Grassmann algebra $(\Lambda (V),{\mbox{\bfseries\slshape g}}%
)$ endowed with the Clifford product is denoted by $\mathcal{C}\ell (V,%
{\mbox{\bfseries\slshape g}})$ or $\mathcal{C}\ell _{p,q}$, the Clifford algebra\footnote{%
If ${\mbox{\bfseries\slshape g}}: V^{\ast }\times V^\ast \rightarrow \mathbb{R}$ we can
analogously also construct the Clifford algebra $\mathcal{C}\ell (V^{\ast },%
\mathtt{g})$, which is of multicovectors, which plays a significant role
when we consider the algebra bundle of multiform fields} associated with $%
V\simeq \mathbb{R}^{p,q},\;p+q=n$. In what follows $\mathbb{R},\mathbb{C}$
denote respectively the real and complex numbers. \ 

Now, restricting to the case where $(p,q)=(1,3)$ we briefly recall the
geometry of Clifford and spin-Clifford bundles. For more details, see e.g. 
\cite{moro}. we denote by $\mathcal{M=}(M,%
{\mbox{\bfseries\slshape g}}%
,\nabla ,\tau _{\mbox{\bfseries\slshape g}},\uparrow )$ the spacetime structure: $M$ denotes a
4-dimensional manifold, $%
{\mbox{\bfseries\slshape g}}%
\in \sec T_{2}^{0}M$ is the metric and in what follows we denote by $g\in
\sec T_{2}^{0}M$ the corresponding metric of the cotangent bundle\footnote{%
If in an arbitrary basis $%
{\mbox{\bfseries\slshape g}}%
=g_{\alpha \beta }dx^{\alpha }\otimes dx^{\beta }$ and $g=g^{\alpha \beta
}\partial _{\alpha }\otimes \partial _{\beta }$ then $g^{\alpha \beta
}g_{\beta \gamma }=\delta _{\gamma }^{\alpha }$.} , $\nabla $ is the
Levi-Civita connection of $%
{\mbox{\bfseries\slshape g}}%
$, $\tau _{\mbox{\bfseries\slshape g}}\in \sec {\displaystyle\Lambda ^{4}}(T^{\ast }M)$ defines a
spacetime orientation and $\uparrow $ refers to an equivalence class of
timelike 1-form fields defining a time orientation. By $F(M)$ we mean the
(principal) bundle of frames, by $\mathbf{P}_{\mathrm{SO}_{1,3}^{e}}(M%
\mathbf{)}$ the orthonormal frame bundle, and $P_{\mathrm{SO}_{1,3}^{e}}(M)$
denotes the orthonormal coframe bundle. We consider $M$ a spin manifold, and
then there exists $\mathbf{P}_{\mathrm{Spin}_{1,3}^{e}}(M\mathbf{)}$ and $P_{%
\mathrm{Spin}_{1,3}^{e}}(M\mathbf{)}$ which are respectively the spin frame
and the spin coframe bundles. We denote by $s:P_{\mathrm{Spin}_{1,3}^{e}}(M%
\mathbf{)\rightarrow }P_{\mathrm{SO}_{1,3}^{e}}(M\mathbf{)}$ the fundamental
mapping present in the definition of $P_{\mathrm{Spin}_{1,3}^{e}}(M\mathbf{)}
$. A spin structure on $M$ consists of a principal fiber bundle $\mathbf{\pi 
}_{s}:P_{\mathrm{Spin}_{1,3}^{e}}(M)\rightarrow M$, with group $\mathrm{Spin}%
_{1,3}^{e}$, and the map 
\begin{equation}
s:P_{\mathrm{Spin}_{1,3}^{e}}(M)\rightarrow P_{\mathrm{SO}_{1,3}^{e}}(M) 
\notag
\end{equation}%
satisfying the following conditions:

(i) $\mathbf{\pi}(s(p))=\mathbf{\pi}_{s}(p),\ \forall p\in P_{\mathrm{Spin}%
_{1,3}^{e}}(M);$ $\pi$ is the projection map of the bundle $P_{\mathrm{SO}%
_{1,3}^{e}}(M)$.

(ii) $s(p \phi)=s(p)\mathrm{Ad}_{\phi},\;\forall p\in P_{\mathrm{Spin}%
_{1,3}^{e}}(M)$ and $\mathrm{Ad}:\mathrm{Spin}_{1,3}^{e}\rightarrow\mathrm{%
Aut}(\mathcal{C}\ell_{1,3}),$ $\mathrm{Ad}_{\phi}:\mathcal{C}\ell_{1,3}\ni
\Xi\mapsto \phi\Xi\phi^{-1}\in\mathcal{C}\ell_{1,3}$ \cite{moro}.

We recall now that sections of $P_{\mathrm{SO}_{1,3}^{e}}(M\mathbf{)}$ are
orthonormal coframes, and that sections of $P_{\mathrm{Spin}_{1,3}^{e}}(M%
\mathbf{)}$ are also orthonormal coframes such that two coframes differing
by a $2\pi$ rotation are distinct and two coframes differing by a $4\pi$
rotation are identified. Next we introduce the Clifford bundle of
differential forms $\mathcal{C\ell (}M,g)$, which is a vector bundle
associated with $P_{\mathrm{Spin}_{1,3}^{e}}(M\mathbf{)}$. Their sections
are sums of non-homogeneous differential forms, which will be called
Clifford fields. We recall that \ $\mathcal{C\ell(}M,g)=P_{\mathrm{SO}%
_{1,3}^{e}}(M)\times _{\mathrm{Ad}^{\prime}}\mathcal{C}\ell_{1,3}$, where $%
\mathcal{C}\ell_{1,3}\simeq$ M(2,${\mathbb{H}})$ is the spacetime algebra 
\cite{rocha1}. Details of the bundle structure are as follows \cite{moro,wal}%
:

(1) Let $\mathbf{\pi}_{c}:\mathcal{C}\ell(M,g)\rightarrow M$ be the
canonical projection of $\mathcal{C}\ell(M,g)$ and let $\{U_{\alpha}\}$ be
an open covering of $M$. There are trivialization mappings $\mathbf{\psi}%
_{i}:\mathbf{\pi}_{c}^{-1}(U_{i})\rightarrow U_{i}\times\mathcal{C}%
\ell_{1,3} $ of the form $\mathbf{\psi}_{i}(p)=(\mathbf{\pi}%
_{c}(p),\psi_{i,x}(p))=(x,\psi_{i,x}(p))$. If $x\in U_{i}\cap U_{j}$ and $%
p\in\mathbf{\pi}_{c}^{-1}(x)$, then 
\begin{equation}
\psi_{i,x}(p)=h_{ij}(x)\psi_{j,x}(p)
\end{equation}
for $h_{ij}(x)\in\mathrm{Aut}(\mathfrak{C}\ell_{1,3})$, where $%
h_{ij}:U_{i}\cap U_{j}\rightarrow\mathrm{Aut}(\mathfrak{C}\ell_{1,3})$ are
the transition mappings of $\mathcal{C}\ell(M,g)$. We recall that every
automorphism of $\mathfrak{C}\ell_{1,3}$ is \textit{inner. }Then, 
\begin{equation}
h_{ij}(x)\psi_{j,x}(p)=a_{ij}(x)\psi_{i,x}(p)a_{ij}(x)^{-1}  \notag
\end{equation}
for some $a_{ij}(x)\in\mathcal{C}\ell_{1,3}^{\star}$, the group of
invertible elements of $\mathcal{C}\ell_{1,3}$.

(2) As it is well known, the group $\mathrm{SO}_{1,3}^{e}$ has a natural
extension in the Clifford algebra $\mathcal{C}\ell _{1,3}$. Indeed, we know
that $\mathcal{C}\ell _{1,3}^{\star }$ (the group of invertible elements of $%
\mathcal{C}\ell _{1,3}$) acts naturally on $\mathcal{C}\ell _{1,3}$ as an
algebra automorphism through its adjoint representation. A set of \emph{lifts%
} of the transition functions of $\mathcal{C}\ell (M,g)$ is a set of
elements $\{a_{ij}\}\subset $ $\mathcal{C}\ell _{1,3}^{\star }$ such that,
if \footnote{%
Recall that $\mathrm{Spin}_{1,3}^{e}=\{\phi \in \mathcal{C}\ell
_{1,3}^{0}:\phi \tilde{\phi}=1\}\simeq \mathrm{SL}(2,\mathbb{C)}$ is the
universal covering group of the restricted Lorentz group $\mathrm{SO}%
_{1,3}^{e}$. Notice that $\mathcal{C}\ell _{1,3}^{0}\simeq \mathcal{C}\ell
_{3,0}\simeq $ M(2, $\mathbb{C})$, the even subalgebra of $\mathcal{C}\ell
_{1,3}$ is the Pauli algebra.} 
\begin{eqnarray}
&&\mathrm{Ad}:\phi \mapsto \mathrm{Ad}_{\phi }  \notag \\
&&\mathrm{Ad}_{\phi }(\Xi )=\phi \Xi \phi ^{-1},\quad \forall \Xi \in 
\mathcal{C}\ell _{1,3},  \notag
\end{eqnarray}%
then $\mathrm{Ad}_{a_{ij}}=h_{ij}$ in all intersections.

(3) Also $\sigma=\mathrm{Ad}|_{\mathrm{Spin}_{1,3}^{e}}$ defines a group
homeomorphism $\sigma:\mathrm{Spin}_{1,3}^{e}\rightarrow\mathrm{SO}%
_{1,3}^{e} $ which is onto with kernel $\mathbb{Z}_{2}$. We have that Ad$%
_{-1}=$ identity, and so $\mathrm{Ad}:\mathrm{Spin}_{1,3}^{e}\rightarrow%
\mathrm{Aut}(\mathcal{C}\ell_{1,3})$ descends to a representation of $%
\mathrm{SO}_{1,3}^{e}$. Let us call $\mathrm{Ad}^{\prime}$ this
representation, i.e., $\mathrm{Ad}^{\prime}:\mathrm{SO}_{1,3}^{e}\rightarrow%
\mathrm{Aut}(\mathcal{C}\ell_{1,3})$. Then we can write $\mathrm{Ad}%
_{\sigma(\phi)}^{\prime}\Xi=\mathrm{Ad}_{\phi}\Xi=\phi\Xi\phi^{-1}$.

(4) It is clear that the structure group of the Clifford bundle $\mathcal{C}%
\ell(M,g)$ is reducible from $\mathrm{Aut}(\mathcal{C}\ell_{1,3})$ to $%
\mathrm{SO}_{1,3}^{e}$. The transition maps of the principal bundle of
oriented Lorentz cotetrads $P_{\mathrm{SO}_{1,3}^{e}}(M)$ can thus be
(through $\mathrm{Ad}^{\prime}$) taken as transition maps for the Clifford
bundle. We then have \cite{lawmi} 
\begin{equation}
\mathcal{C}\ell(M,g)=P_{\mathrm{SO}_{1,3}^{e}}(M)\times _{\mathrm{Ad}%
^{\prime}}\mathcal{C}\ell_{1,3},  \notag
\end{equation}
i.e., the Clifford bundle is a vector bundle associated with the principal
bundle $P_{\mathrm{SO}_{1,3}^{e}}(M)$ of orthonormal Lorentz coframes.

Recall that $\mathcal{C}\!\ell (T_{x}^{\ast }M,g_{x})$ is also a vector
space over $\mathbb{R}$ which is isomorphic to the exterior algebra $\Lambda
(T_{x}^{\ast }M)$ of the cotangent space and $\Lambda (T_{x}^{\ast
}M)=\bigoplus\nolimits_{k=0}^{4}\Lambda {}^{k}(T_{x}^{\ast }M)$, where $%
\Lambda ^{k}(T_{x}^{\ast }M)$ is the $\binom{4}{k}$-dimensional space of $k$%
-forms over a point $x$ on $M$. There is a natural embedding \ $\Lambda
(T^{\ast }M)\hookrightarrow $ $\mathcal{C}\ell (M,g)$ \cite{lawmi} and
sections of $\mathcal{C}\!\ell (M,g)$ --- Clifford fields --- can be
represented as a sum of non-homogeneous differential forms. Let $%
\{e_{a}\}\in \sec \mathbf{P}_{\mathrm{SO}_{1,3}^{e}}(M)$ (the orthonormal
frame bundle) be a tetrad basis for $TU\subset TM$ (given an open set $%
U\subset M$). Moreover, let $\{\vartheta ^{a}\}\in \sec P_{\mathrm{SO}%
_{1,3}^{e}}(M)$. Then, for each ${a}=0,1,2,3$, ${\vartheta }^{a}\in \sec
\Lambda ^{1}(T^{\ast }M)\hookrightarrow \sec \mathcal{C}\!\ell (M,g$). We
recall next the crucial result \cite{moro} that in a spin manifold we have: 
\begin{equation}
\mathcal{C}\ell (M,g)=P_{\mathrm{Spin}_{1,3}^{e}}(M)\times _{\mathrm{Ad}}%
\mathcal{C}\ell _{1,3}.  \notag
\end{equation}%
Spinor fields are sections of vector bundles associated with the principal
bundle of spinor coframes. The well known Dirac spinor fields are sections
of the bundle 
\begin{equation}
S_{c}(M,g)=P_{\mathrm{Spin}_{1,3}^{e}}(M)\times _{\mu _{c}}\mathbb{C}^{4}, 
\notag
\end{equation}%
with $\mu _{c}$ the $D^{(1/2,0)}\oplus D^{(0,1/2)}$ representation of $%
\mathrm{Spin}_{1,3}^{e}\cong \mathrm{SL}(2,\mathbb{C})$ in $\mathrm{End}(%
\mathbb{C}^{4})$ \cite{choquet}.

\section{Bilinear Covariants and ELKO spinor fields}

This Section is devoted to recall the bilinear covariants, using the
programme introduced in \cite{meu}, which we briefly recall here. In this
article all spinor fields live in Minkowski spacetime $(M,\eta ,D,\tau
_{\eta },\uparrow )$. The manifold $M$ $\simeq \mathbb{R}^{4}$, $\eta $
denotes a constant metric, where $\eta (\partial /\partial x^{\mu },\partial
/\partial x^{\nu })=\eta _{\mu \nu }=\mathrm{diag}(1,-1,-1,-1)$, $D$ denotes
the Levi-Civita connection associated with $\eta $, $M$ is oriented by the
4-volume element $\tau _{\eta }$ and time-oriented by $\uparrow $. Here $%
\{x^{\mu }\}$ denotes global coordinates in the Einstein-Poincar\'{e} gauge,
naturally adapted to an inertial reference frame $\mathbf{e}_{0}=\partial
/\partial x^{0}$. Let $\mathbf{e}_{i}=\partial /\partial x^{i}$, $i=1,2,3$.
Also, $\{\mathbf{e}_{\mu }\}$ is a section of the frame bundle $\mathbf{P}_{%
\mathrm{SO}_{1,3}^{e}}(M)$ and $\{\mathbf{e}^{\mu }\}$ is its reciprocal
frame satisfying $\eta (\mathbf{e}^{\mu },\mathbf{e}_{\nu }):=\mathbf{e}%
^{\mu }\cdot \mathbf{e}_{\nu }=\delta _{\nu }^{\mu }$. \ Let moreover be $%
\{\theta ^{\mu }\}$ and $\{\theta _{\mu }\}$ be respectively the dual bases
\ of \ $\{\mathbf{e}_{\mu }\}$ and $\{\mathbf{e}^{\mu }\}$. Classical spinor
fields carrying a $D^{(1/2,0)}\oplus D^{(0,1/2)}$ representation of SL$(2,%
\mathbb{C)\simeq }\;\,\mathrm{Spin}_{1,3}^{e}$ are sections of the vector
bundle $\mathbf{P}_{\mathrm{Spin}_{1,3}^{e}}(M)\times _{\rho }\mathbb{C}%
^{4}, $ where $\rho $ stands for the $D^{(1/2,0)}\oplus D^{(0,1/2)}$
representation of SL$(2,\mathbb{C)\simeq }\;\,\mathrm{Spin}_{1,3}^{e}$ in $%
\mathbb{C}^{4}$. In addition, classical spinor fields carrying a $%
D^{(1/2,0)} $ or $D^{(0,1/2)})$ representation of SL$(2,\mathbb{C)\simeq }%
\;\,\mathrm{Spin}_{1,3}^{e}$ are sections of the vector bundle $\mathbf{P}_{%
\mathrm{Spin}_{1,3}^{e}}(M)\times _{\rho ^{\prime }}\mathbb{C}^{2},$ where $%
\rho ^{\prime }$ stands for the $D^{(1/2,0)}$ or the $D^{(0,1/2)}$
representation of SL$(2,\mathbb{C)\simeq }\;\,\mathrm{Spin}_{1,3}^{e}$ in $%
\mathbb{C}^{2}$. Given a spinor field $\psi $ $\in \sec \mathbf{P}_{\mathrm{%
Spin}_{1,3}^{e}}(M)\times _{\rho }\mathbb{C}^{4}$ the bilinear covariants
may be taken as the following sections of the exterior algebra bundle of 
\textit{multiform} fields $\bigwedge T^{\ast }M\hookrightarrow \mathcal{C}%
\ell (M,\mathtt{\eta })$ \ ( where $\mathcal{C}\ell (M,\mathtt{\eta })$ is
the Clifford bundle of multiform fields and $\mathtt{\eta }$, of course,
denotes the metric of the cotangent bundle \cite{moro}): 
\begin{align}
\sigma & =\psi ^{\dagger }\gamma _{0}\psi ,\quad \mathbf{J}=J_{\mu }\theta
^{\mu }=\psi ^{\dagger }\gamma _{0}\gamma _{\mu }\psi \theta ^{\mu },\quad 
\mathbf{S}=S_{\mu \nu }\theta ^{\mu \nu }=\frac{1}{2}\psi ^{\dagger }\gamma
_{0}i\gamma _{\mu \nu }\psi \theta ^{\mu }\wedge \theta ^{\nu },  \notag \\
\mathbf{K}& =K_{\mu }\theta ^{\mu }=\psi ^{\dagger }\gamma _{0}i\gamma
_{0123}\gamma _{\mu }\psi \theta ^{\mu },\quad \omega =-\psi ^{\dagger
}\gamma _{0}\gamma _{0123}\psi ,  \label{fierz}
\end{align}%
The set $\{\gamma _{\mu }\}$ refers to the Dirac matrices in chiral
representation (see Eq.(\ref{dirac matrices})). Also $\{\mathbf{1}%
_{4},\gamma _{\mu },\gamma _{\mu }\gamma _{\nu },\gamma _{\mu }\gamma _{\nu
}\gamma _{\rho },\gamma _{0}\gamma _{1}\gamma _{2}\gamma _{3}\}$ ($\mu ,\nu
,\rho =0,1,2,3$, and $\mu <\nu <\rho $) is a basis for $\mathbb{C}(4)$
satisfying \cite{lou1} $\gamma _{\mu }\gamma _{\nu }+\gamma _{\nu }\gamma
_{\mu }=2\eta _{\mu \nu }\mathbf{1}_{4}$ and the Clifford product is denoted
by juxtaposition. More details on notations can be found in \cite{moro,rod}.

In the case of the electron, described by Dirac spinor fields (classes 1, 2
and 3 below), $\mathbf{J}$ is a future-oriented timelike current vector
which gives the current of probability, the bivector $\mathbf{S}$ is
associated with the distribution of intrinsic angular momentum, and the
spacelike vector $\mathbf{K}$ is associated with the direction of the
electron spin. For a detailed discussion concerning such entities, their
relationships and physical interpretation, and generalizations, see, e.g., 
\cite{cra,lou1,lou2,holl,hol,doran,dor2}.

The bilinear covariants satisfy the Fierz identities \cite%
{cra,lou1,lou2,holl,hol} 
\begin{equation}
\mathbf{J}^{2}=\omega^{2}+\sigma^{2},\quad\mathbf{K}^{2}=-\mathbf{J}%
^{2},\quad\mathbf{J}\llcorner\mathbf{K}=0,\quad\mathbf{J}\wedge\mathbf{K}%
=-(\omega+\sigma\gamma_{0123})\mathbf{S}.  \label{fi}
\end{equation}
\noindent A spinor field such that \emph{not both} $\omega$ and $\sigma$ are
null is said to be regular. When $\omega=0=\sigma$, a spinor field is said
to be \textit{singular}.

Lounesto spinor field classification is given by the following spinor field
classes \cite{lou1,lou2}, where in the first three classes it is implicit
that $\mathbf{J}$\textbf{, }$\mathbf{K}$\textbf{, }$\mathbf{S}$ $\neq0$:

\begin{itemize}
\item[1)] $\sigma\neq0,\;\;\; \omega\neq0$.

\item[2)] $\sigma\neq0,\;\;\; \omega= 0$.\label{dirac1}

\item[3)] $\sigma= 0, \;\;\;\omega\neq0$.\label{dirac2}

\item[4)] $\sigma= 0 = \omega, \;\;\;\mathbf{K}\neq0,\;\;\; \mathbf{S}\neq0$.%
\label{tipo4}

\item[5)] $\sigma= 0 = \omega, \;\;\;\mathbf{K}= 0, \;\;\;\mathbf{S}\neq0$.%
\label{elko1}

\item[6)] $\sigma= 0 = \omega, \;\;\; \mathbf{K}\neq0, \;\;\; \mathbf{S} = 0$%
.
\end{itemize}

\noindent The current density $\mathbf{J}$ is always non-zero. Types-(1),
-(2), and -(3) spinor fields are denominated \textit{Dirac spinor fields}
for spin-1/2 particles and types-(4), -(5), and -(6) are respectively called 
\textit{flag-dipole}, \textit{flagpole}\footnote{%
Such spinor fields are constructed by a null 1-form field current and an
also null 2-form field angular momentum, the ``flag'' \cite{koso}.} and 
\textit{Weyl spinor fields}. Majorana spinor fields are a particular case of
a type-(5) spinor field. It is worthwhile to point out a peculiar feature of
types-(4), -(5) and -(6) spinor fields: although $\mathbf{J}$ is always
non-zero, $\mathbf{J}^{2}=-\mathbf{K}^{2}=0$. It shall be seen below that
the bilinear covariants related to an ELKO spinor field, satisfy $%
\sigma=0=\omega,\;\;\mathbf{K}=0,\;\;\mathbf{S}\neq0$ and $\mathbf{J}^{2}=0$%
. Since Lounesto proved that there are \textit{no} other classes based on
distinctions among bilinear covariants, ELKO spinor fields must belong to
one of the disjoint six classes. 

Types-(1), -(2) and -(3) Dirac spinor fields (DSFs) have different algebraic
and geometrical characters, and we would like to emphasize the main
differing points. For more details, see e.g. \cite{lou1,lou2}. Recall that
if the quantities $P=\sigma +\mathbf{J}+\gamma _{0123}\omega $ and $Q=%
\mathbf{S}+\mathbf{K}\gamma _{0123}$ are defined \cite{lou1,lou2}, in
type-(1) DSF we have $P=-(\omega +\sigma \gamma _{0123})^{-1}\mathbf{K}Q$
and also $\psi =-i(\omega +\sigma \gamma _{0123})^{-1}\psi $. In type-(2)
DSF, $P$ is a multiple of $\frac{1}{2\sigma }(\sigma +\mathbf{J})$ and looks
like a proper energy projection operator, commuting with the spin projector
operator given by $\frac{1}{2}(1-i\gamma _{0123}\mathbf{K}/\sigma )$. Also, $%
P=\gamma _{0123}\mathbf{K}Q/\sigma $. Further, in type-(3) DSF, $P^{2}=0$
and $P=\mathbf{K}Q/\omega $. The introduction of the spin-Clifford bundle
makes it possible to consider all the geometric and algebraic objects ---
the Clifford bundle, spinor fields, differential form fields, operators and
Clifford fields --- as being elements of an unique unified formalism. It is
well known that spinor fields have three different, although equivalent,
definitions: the operator, the classical and the algebraic one. In
particular, the operatorial definition allows us to factor --- up to sign
--- the DSF $\psi $ as $\psi =(\sigma +\omega \gamma _{0123})^{-1/2}R$,
where $R\in $ Spin$_{1,3}^{e}(M)\hookrightarrow \mathcal{C}\ell (M,\mathtt{%
\eta })$. Finally, to a Weyl spinor field $\xi $ (type-(6)) with bilinear
covariants \textbf{J} and \textbf{K}, two Majorana spinor fields $\psi _{\pm
}=\frac{1}{2}(\xi +C(\xi ))$ can be associated, where $C$ denotes the charge
conjugation operator. Penrose flagpoles are implicitly defined by the
equation $\sigma +\mathbf{J}+i\mathbf{S}-i\gamma _{0123}\mathbf{K}+\gamma
_{0123}\omega =\frac{1}{2}(\mathbf{J}\mp i\mathbf{S}\gamma _{0123})$ \cite%
{lou1,lou2}. For a physically useful discussion regarding the disjoint
classes -(5) and -(6) see, e.g., \cite{plaga}. The fact that two Majorana
spinor fields $\psi _{\pm }$ can be written in terms of a Weyl type-(6)
spinor field $\psi _{\pm }=\frac{1}{2}(\xi +C(\xi ))$, is an `accident' when
the (Lorentzian) spacetime has $n=4$ --- the present case --- or $n=6$
dimensions. The more general assertion concerns the property that two
Majorana, and more generally ELKO spinor fields $\psi _{\pm }$ can be
written in terms of a \emph{pure spinor} field \cite{bud,cru} --- hereon
denoted by $\mathfrak{u}$ --- as $\psi _{\pm }=\frac{1}{2}(\mathfrak{u}+C(%
\mathfrak{u}))$. It is well known that Weyl spinor fields are pure spinor
fields when $n=4$ and $n=6$. When the complexification $\mathbb{C}\otimes 
\mathbb{R}^{1,3}$ of $\mathbb{R}^{1,3}$ is considered, one can consider a
maximal totally isotropic subspace $N$ of $\mathbb{C}\otimes \mathbb{R}%
^{1,3} $, by the Witt decomposition, where $\dim _{\mathbb{C}}N=2$. Pure
spinors are defined by the property $x\mathfrak{u}=0$ for all $x\in N\subset 
\mathbb{C}\otimes \mathbb{R}^{1,3}$ \cite{cru}. In this context, Penrose
flags can be defined by the expression Re$(i\mathfrak{u}\tilde{\mathfrak{u}}%
) $ \cite{benn}.

Now, the algebraic and formal properties of ELKO spinor fields, as defined
in \cite{allu,alu2,alu3,meu}, are briefly explored. An ELKO $\Psi$
corresponding to a plane wave with momentum $p=(p^{0},\mathbf{p)}$ can be
written, without loss of generality, as $\Psi(p)=\lambda(\mathbf{p}) e^{-i{%
p\cdot x}}$ (or $\Psi(p)=\lambda(\mathbf{p}) e^{i{p\cdot x}}$) where 
\begin{equation}
\lambda(\mathbf{p})=\binom{i\Theta\phi_{L}^{\ast}(\mathbf{p})}{\phi_{L}(%
\mathbf{p})},  \label{1}
\end{equation}
\noindent and the Wigner's spin-1/2 time reversal operator $\Theta$
satisfies $\Theta \mathfrak{J}\Theta^{-1}=-\mathfrak{J}^{\ast}$, where ${%
\mathfrak{J}}$ denotes the generators of rotations \cite{allu}. Hereon, as
in \cite{allu}, the Weyl representation of $\gamma^{\mu}$ is used, i.e., 
\begin{equation}
\gamma_{0}=\gamma^{0}=%
\begin{pmatrix}
{\mathbb{O}} & {\mathbb{I}} \\ 
{\mathbb{I}} & {\mathbb{O}}%
\end{pmatrix}
,\quad-\gamma_{k}=\gamma^{k}=%
\begin{pmatrix}
{\mathbb{O}} & -\sigma_{k} \\ 
\sigma_{k} & {\mathbb{O}}%
\end{pmatrix}
,  \label{dirac matrices}
\end{equation}
\noindent where 
\begin{equation}
{\mathbb{I}}= 
\begin{pmatrix}
1 & 0 \\ 
0 & 1%
\end{pmatrix}
,\quad {\mathbb{O}}=%
\begin{pmatrix}
0 & 0 \\ 
0 & 0%
\end{pmatrix}
,\quad \sigma_{1}=%
\begin{pmatrix}
0 & 1 \\ 
1 & 0%
\end{pmatrix}
,\quad\sigma_{2}=%
\begin{pmatrix}
0 & -i \\ 
i & 0%
\end{pmatrix}
,\quad\sigma_{3}=%
\begin{pmatrix}
1 & 0 \\ 
0 & -1%
\end{pmatrix}%
,  \notag
\end{equation}
\noindent $\sigma_i$ are the Pauli matrices. Also, 
\begin{equation}
\gamma^{5}=i\gamma^{0}\gamma^{1}\gamma^{2}\gamma^{3}=i\gamma^{0123}=-i%
\gamma_{0123} = 
\begin{pmatrix}
{\mathbb{I}} & {\mathbb{O}} \\ 
{\mathbb{O}} & -{\mathbb{I}}%
\end{pmatrix}%
.  \notag
\end{equation}
\noindent ELKO spinor fields are eigenspinors of the charge conjugation
operator $C$, i.e., $C\lambda(\mathbf{{p})=\pm \lambda(p)}$, for 
\begin{equation}
C=%
\begin{pmatrix}
{\mathbb{O}} & i\Theta \\ 
-i\Theta & {\mathbb{O}}%
\end{pmatrix}
K .  \label{conj}
\end{equation}
The operator $K$ is responsible for the $\mathbb{C}$-conjugation of Weyl
spinor fields appearing on the right. The plus sign stands for \textit{%
self-conjugate} spinors, $\lambda^{S}(\mathbf{p})$, while the minus yields 
\textit{anti self-conjugate} spinors, $\lambda^{A}(\mathbf{p})$. Explicitly,
the complete form of ELKO spinor fields can be found by solving the equation
of helicity $(\sigma\cdot\widehat{\mathbf{p}})\phi^{\pm}=\pm \phi^{\pm}$ in
the rest frame and subsequently make a boost, to recover the result for any $%
\mathbf{p}$ \cite{allu}. Here $\widehat{\mathbf{p}}:=\mathbf{p}/\|\mathbf{p}%
\|$. The four spinor fields are given 
\begin{equation}
\lambda^{S/A}_{\{\mp,\pm \}}({\ p})=\sqrt{\frac{E+m}{2m}}\Bigg(1\mp \frac{%
\mathbf{p}}{E+m}\Bigg)\lambda^{S/A}_{\{\mp,\pm \}}(\mathbf{{0}) ,}
\label{form}
\end{equation}
where 
\begin{equation}
\overset{\neg}{\lambda}^{S/A}_{\{\mp,\pm \}}(\mathbf{p})=\pm i \Big[ %
\lambda^{S/A}_{\{\pm,\mp \}}(\mathbf{p})\Big]^{\dag}\gamma^{0} .
\label{dual}
\end{equation}
Note that, since $\Theta[\phi^{\pm}(\mathbf{0})]^{*}$ and $%
\phi^{\pm}({\mathbf{{0}}})$ have opposite helicities, ELKO cannot be an
eigenspinor field of the helicity operator, and indeed carries both
helicities. In order to guarantee an invariant real norm, as well as
positive definite norm for two ELKO spinor fields, and negative definite
norm for the other two, the ELKO dual is given by 
\begin{equation}
\overset{\neg}{\lambda}^{S/A}_{\{\mp,\pm \}}(\mathbf{p})=\pm i \Big[ %
\lambda^{S/A}_{\{\pm,\mp \}}(\mathbf{p})\Big]^{\dag}\gamma^{0} .
\label{dual1}
\end{equation}

Omitting the subindex of the spinor field $\phi _{L}(\mathbf{p})$, which is
denoted hereon by $\phi $, the left-handed spinor field $\phi _{L}(\mathbf{p}%
)$ can be represented by 
\begin{equation}
\phi =\binom{\alpha (\mathbf{p})}{\beta (\mathbf{p})},\quad \alpha (\mathbf{p%
}),\beta (\mathbf{p})\in \mathbb{C}.  \label{0}
\end{equation}%
\noindent \noindent Now using Eqs.(\ref{fierz}) it is possible to \emph{%
calculate} explicitly the bilinear covariants for ELKO spinor fields%
\footnote{%
All the details are presented in \cite{meu}.}: 
\begin{align}
\mathring{\sigma}& =\lambda ^{\dagger }\gamma _{0}\lambda =0,\qquad 
\mathring{\omega}=-\lambda ^{\dagger }\gamma _{0}\gamma _{0123}\lambda =0 
\notag \\
\mathbf{\mathring{J}}& =\mathring{J}_{\mu }\theta ^{\mu }=\lambda ^{\dagger
}\gamma _{0}\gamma _{\mu }\lambda \theta ^{\mu }\neq 0  \notag \\
\mathbf{\mathring{K}}& =\mathring{K}_{\mu }\theta ^{\mu }=\lambda ^{\dagger
}i\gamma _{123}\gamma _{\mu }\lambda \theta ^{\mu }=0,  \notag \\
\mathbf{\mathring{S}}& =\frac{1}{2}\mathring{S}_{\mu \nu }\theta ^{\mu \nu }=%
\frac{1}{2}\lambda ^{\dagger }\gamma _{0}i\gamma _{\mu \nu }\lambda \theta
^{\mu \nu }\neq 0.  \notag
\end{align}%
\noindent Indeed, since the relations 
\begin{eqnarray}
\sigma _{1}\phi &=&%
\begin{pmatrix}
0 & 1 \\ 
1 & 0%
\end{pmatrix}%
\binom{\alpha }{\beta }=\binom{\beta }{\alpha },\quad \sigma _{2}\phi =%
\begin{pmatrix}
0 & -i \\ 
i & 0%
\end{pmatrix}%
\binom{\alpha }{\beta }=\binom{-i\beta }{i\alpha },  \notag \\
\sigma _{3}\phi &=&%
\begin{pmatrix}
1 & 0 \\ 
0 & -1%
\end{pmatrix}%
\binom{\alpha }{\beta }=\binom{\alpha }{-\beta }.  \notag
\end{eqnarray}%
\noindent hold, Eq.(\ref{1}) gives $\psi ^{\dagger }=[(\sigma _{2}\phi
^{\ast })^{\dagger },\phi ^{\dagger }]=[(i\beta ,-i\alpha ),(\alpha ^{\ast
},\beta ^{\ast })],$ and thus 
\begin{eqnarray}
\mathring{\sigma} &=&\psi ^{\dagger }\gamma _{0}\psi  \notag \\
&=&[(i\beta ,-i\alpha ),(\alpha ^{\ast },\beta ^{\ast })]\binom{\binom{%
\alpha }{\beta }}{\binom{-i\beta ^{\ast }}{i\alpha ^{\ast }}}  \notag \\
&=&i\beta \alpha -i\alpha \beta -i\alpha ^{\ast }\beta ^{\ast }+i\beta
^{\ast }\alpha ^{\ast }  \notag \\
&=&0,  \notag \\
&&  \notag \\
\mathring{\omega} &=&-\psi ^{\dagger }\gamma _{123}\psi =[(i\beta ,-i\alpha
),(\alpha ^{\ast },\beta ^{\ast })]\binom{\binom{i\alpha }{i\beta }}{\binom{%
-\beta ^{\ast }}{\alpha ^{\ast }}}  \notag \\
&=&0,  \notag \\
&&  \notag \\
\mathring{\bf J} &=&\mathring{J}_{\mu }\gamma ^{\mu }=\psi ^{\dagger }\gamma
_{0}\gamma _{\mu }\psi \gamma ^{\mu }  \notag \\
&=&\psi ^{\dagger }\gamma _{0}\gamma _{1}\psi \gamma ^{1}+\psi ^{\dagger
}\gamma _{0}\gamma _{2}\psi \gamma ^{2}+\psi ^{\dagger }\gamma _{0}\gamma
_{3}\psi \gamma ^{3}-\psi ^{\dagger }\psi \gamma ^{0}  \notag \\
&=&\psi ^{\dagger }%
\begin{pmatrix}
0 & 1 \\ 
1 & 0%
\end{pmatrix}%
\begin{pmatrix}
0 & -\sigma_1 \\ 
\sigma_1 & 0%
\end{pmatrix}%
\psi \gamma ^{1}+\psi ^{\dagger }%
\begin{pmatrix}
0 & 1 \\ 
1 & 0%
\end{pmatrix}%
\begin{pmatrix}
0 & -\sigma_2 \\ 
\sigma_2 & 0%
\end{pmatrix}%
\psi \gamma ^{2}  \notag \\
&&+\psi ^{\dagger }%
\begin{pmatrix}
0 & 1 \\ 
1 & 0%
\end{pmatrix}%
\begin{pmatrix}
0 & -\sigma_3 \\ 
\sigma_3 & 0%
\end{pmatrix}%
\psi \gamma ^{3}-\psi ^{\dagger }%
\begin{pmatrix}
0 & i \\ 
-i & 0%
\end{pmatrix}%
\psi \gamma ^{0}  \notag \\
&=&\psi ^{\dagger }\binom{i\sigma _{3}\phi ^{\ast }}{-\sigma _{1}\phi }%
\gamma ^{1}+\psi ^{\dagger }\binom{-\phi ^{\ast }}{-\sigma _{2}\phi }\gamma
^{2}-\sigma ^{\dagger }\binom{-i\sigma _{1}\phi ^{\ast }}{-\sigma _{3}\phi }%
+\psi ^{\dagger }\psi \gamma ^{0}  \notag \\
&=&[(i\beta ,-i\alpha ),(\alpha ^{\ast },\beta ^{\ast })]\binom{\binom{%
i\alpha ^{\ast }}{-i\beta ^{\ast }}}{\binom{-\beta }{-\alpha }}\gamma
^{1}+[(i\beta ,-i\alpha ),(\alpha ^{\ast },\beta ^{\ast })]\binom{\binom{%
-\alpha ^{\ast }}{-\beta ^{\ast }}}{\binom{i\beta }{-i\alpha }}\gamma ^{2}+ 
\notag \\
&&[(i\beta ,-i\alpha ),(\alpha ^{\ast },\beta ^{\ast })]\binom{\binom{%
-i\beta ^{\ast }}{-i\alpha ^{\ast }}}{\binom{-\alpha }{\beta }}\gamma
^{3}+[(i\beta ,-i\alpha ),(\alpha ^{\ast },\beta ^{\ast })]\binom{\binom{%
-i\beta ^{\ast }}{i\alpha ^{\ast }}}{\binom{\alpha }{\beta }}\gamma ^{0} 
\notag \\
&=&2(\alpha \beta ^{\ast }+\alpha ^{\ast }\beta )\gamma ^{1}+2i(\alpha
^{\ast }\beta -\alpha \beta ^{\ast })\gamma ^{2}+2(\beta \beta ^{\ast
}-\alpha \alpha ^{\ast })\gamma ^{3}  \notag \\
&&\quad +2(\alpha \alpha ^{\ast }+\beta \beta ^{\ast })\gamma ^{0}  \notag \\
&\neq &0,  \notag \\
&&  \notag \\
\mathring{\bf K} &=&\mathring{K}_{\mu }\gamma ^{\mu }=\psi ^{\dagger
}i\gamma _{123}\gamma _{\mu }\psi \gamma ^{\mu }  \notag \\
&=&i\psi ^{\dagger }%
\begin{pmatrix}
-i\sigma_1 & 0 \\ 
0 & -i\sigma_1%
\end{pmatrix}%
\psi \gamma ^{1}+i\psi ^{\dagger }%
\begin{pmatrix}
-i\sigma_2 & 0 \\ 
0 & -i\sigma_2%
\end{pmatrix}%
\psi \gamma ^{2}  \notag \\
&&\quad +i\psi ^{\dagger }%
\begin{pmatrix}
-i\sigma_3 & 0 \\ 
0 & -i\sigma_3%
\end{pmatrix}%
\psi \gamma ^{3}+i\psi ^{\dagger }%
\begin{pmatrix}
1 & 0 \\ 
0 & -1%
\end{pmatrix}%
\psi \gamma ^{0}  \notag \\
&=&\psi ^{\dagger }\binom{i\sigma _{3}\phi ^{\ast }}{\sigma _{1}\phi }\gamma
^{1}-\psi ^{\dagger }\binom{\phi ^{\ast }}{-\sigma _{2}\phi }\gamma
^{2}-\psi ^{\dagger }\binom{i\sigma _{1}\phi ^{\ast }}{-\sigma _{3}\phi }%
\gamma ^{3}+\psi ^{\dagger }\binom{\sigma _{2}\phi ^{\ast }}{-\phi }\gamma
^{0}  \notag \\
&=&[(i\beta ,-i\alpha ),(\alpha ^{\ast },\beta ^{\ast })]\binom{\binom{%
\alpha ^{\ast }}{-\beta ^{\ast }}}{\binom{-i\beta }{-i\alpha }}\gamma
^{1}+[(i\beta ,-i\alpha ),(\alpha ^{\ast },\beta ^{\ast })]\binom{\binom{%
i\alpha ^{\ast }}{i\beta ^{\ast }}}{\binom{-\beta }{\alpha }}\gamma ^{2} 
\notag \\
&&+[(i\beta ,-i\alpha ),(\alpha ^{\ast },\beta ^{\ast })]\binom{\binom{%
-\beta ^{\ast }}{-\alpha ^{\ast }}}{\binom{-i\alpha }{i\beta }}\gamma
^{3}+[(i\beta ,-i\alpha ),(\alpha ^{\ast },\beta ^{\ast })]\binom{\binom{%
-i\beta ^{\ast }}{i\alpha ^{\ast }}}{\binom{\alpha }{\beta }}\gamma ^{0} 
\notag \\
&=&0.  \notag
\end{eqnarray}%
Finally, the value for $\mathring{\bf S}$ is now computed: 
\begin{eqnarray}
\mathring{\bf S} &=&\frac{1}{2}\mathring{S}_{\mu \nu }\gamma ^{\mu \nu }=%
\frac{1}{2}\psi ^{\dagger }\gamma _{0}i\gamma _{\mu \nu }\psi \gamma ^{\mu
\nu }  \notag \\
&=&\frac{i}{2}(\psi ^{\dagger }\gamma _{1}\psi \gamma ^{01}+\psi ^{\dagger
}\gamma _{2}\psi \gamma ^{02}+\psi ^{\dagger }\gamma _{3}\psi \gamma
^{03}+\psi ^{\dagger }\gamma _{012}\psi \gamma ^{12}+\psi ^{\dagger }\gamma
_{013}\psi \gamma ^{13}+\psi ^{\dagger }\gamma _{023}\gamma ^{23})  \notag \\
&=&\frac{i}{2}\left( \psi ^{\dagger }\binom{-\sigma _{1}\phi ^{\ast }}{%
i\sigma _{3}\phi }\gamma ^{01}-\psi ^{\dagger }\binom{\sigma _{2}\phi }{\phi
^{\ast }}\gamma ^{02}-\psi ^{\dagger }\binom{\sigma _{3}\phi }{i\sigma
_{1}\phi ^{\ast }}\gamma ^{03}-\psi ^{\dagger }\binom{i\sigma _{3}\phi }{%
\sigma _{1}\phi ^{\ast }}\gamma ^{12}\right)  \notag \\
&&\quad +\frac{i}{2}\psi ^{\dagger }\binom{-i\sigma _{1}\phi }{\sigma
_{3}\phi ^{\ast }}\gamma ^{23}-\frac{1}{2}\psi ^{\dagger }\binom{\sigma
_{2}\phi }{-\phi ^{\ast }}\gamma ^{13}  \notag \\
&=&\frac{i}{2}\{[(i\beta ,-i\alpha ),(\alpha ^{\ast },\beta ^{\ast })]\binom{%
\binom{-\beta ^{\ast }}{-\alpha ^{\ast }}}{\binom{i\alpha }{-i\beta }}\gamma
^{01}+[(i\beta ,-i\alpha ),(\alpha ^{\ast },\beta ^{\ast })]\binom{\binom{%
i\beta }{-i\alpha }}{\binom{-\alpha ^{\ast }}{-\beta ^{\ast }}}\gamma ^{02} 
\notag \\
&&+[(i\beta ,-i\alpha ),(\alpha ^{\ast },\beta ^{\ast })]\binom{\binom{%
-\alpha }{\beta }}{\binom{-i\beta ^{\ast }}{-i\alpha ^{\ast }}}\gamma
^{03}+[(i\beta ,-i\alpha ),(\alpha ^{\ast },\beta ^{\ast })]\binom{\binom{%
-i\alpha }{i\beta }}{\binom{-\beta ^{\ast }}{-\alpha ^{\ast }}}\gamma ^{12} 
\notag \\
&&+[(i\beta ,-i\alpha ),(\alpha ^{\ast },\beta ^{\ast })]\binom{\binom{%
-i\beta }{i\alpha }}{\binom{-\alpha ^{\ast }}{-\beta ^{\ast }}}\gamma
^{13}+[(i\beta ,-i\alpha ),(\alpha ^{\ast },\beta ^{\ast })]\binom{\binom{%
-i\beta }{-i\alpha }}{\binom{\alpha ^{\ast }}{-\beta ^{\ast }}}\gamma ^{23}\}
\notag \\
&=&\frac{i}{2}((\alpha ^{\ast })^{2}+(\beta ^{\ast })^{2}-\beta ^{2}-\alpha
^{2})\gamma ^{02}+\frac{1}{2}((\alpha ^{\ast })^{2}+(\beta ^{\ast
})^{2}+\beta ^{2}+\alpha ^{2})\gamma ^{31}  \notag \\
&&\quad +\frac{1}{2}((\beta ^{\ast })^{2}+\beta ^{2}-(\alpha ^{\ast
})^{2}-\alpha ^{2})\gamma ^{01}+\frac{i}{2}(-\beta ^{2}-\alpha ^{2}+(\alpha
^{\ast })^{2}+(\beta ^{\ast })^{2})\gamma ^{02}  \notag \\
&&\quad +(\alpha \beta +\alpha ^{\ast }\beta ^{\ast })\gamma ^{03}+\frac{i}{2%
}(\alpha \beta -\alpha ^{\ast }\beta ^{\ast })\gamma ^{12}+\frac{i}{2}(\beta
^{2}-\alpha ^{2}+(\alpha ^{\ast })^{2}-({\beta ^{\ast })}^{2})\gamma ^{23}. 
\notag \\
&\neq &0.  \notag
\end{eqnarray}%
From these formul\ae \thinspace\ it is trivially seen that that $\mathring{%
\bf{J}}\lrcorner \mathring{\bf{K}}=0.$ The relations above give $\mathring{%
\mathbf{J}}^{2}=0,$ and it is immediate that all Fierz identities introduced
by the formul\ae \thinspace\ in Eqs.(\ref{fi}) are trivially satisfied.

It is useful to choose $i\Theta=\sigma_{2}$, as in \cite{allu}, in such a
way that it is possible to express 
\begin{equation}
\lambda=\binom{\sigma_{2}\phi_{L}^{\ast}(\mathbf{{p})}}{\phi_{L}(\mathbf{{p})%
}}.  \label{01}
\end{equation}

Now, any flagpole spinor field is an eigenspinor field of the charge
conjugation operator \cite{lou1,lou2}, which explicit action on a spinor $%
\psi$ is given by $\mathcal{C}\psi=-\gamma ^{2}\psi^{\ast}$. Indeed using
Eq.(\ref{01}) it follows that 
\begin{eqnarray}
-\gamma^{2}\lambda^{\ast} &=&\binom{\sigma_{2}\phi^{\ast}}{%
-\sigma_{2}\sigma_{2}^{\ast}\phi}  \notag \\
&=&\lambda.  \notag
\end{eqnarray}
\noindent

Once the definition of ELKO spinor fields is recalled, we return to the
previous discussion about Penrose flagpoles. Here we extend the definition
of the Penrose poles, and we can prove that they are given in terms of an
ELKO spinor field by the expression $\frac{1}{2}\langle\lambda(\widetilde{%
\gamma_{0123}\lambda})\rangle_1$, and further, Penrose flags $F$ can also be
written in terms of ELKO, as $F = \frac{1}{2}\langle\lambda(\widetilde{%
\gamma_{0123}\lambda})\rangle_2$. This assertion can be demonstrated
following a reasoning analogous to the one exposed in \cite{benn,chev}.

\section{Classical, ideal and operator spinors in the spacetime algebra}

\label{scle} Given an orthonormal basis $\{\mathbf{e}_{\mu }\}$ in $\mathbb{R%
}^{1,3}$ an arbitrary element of ${\mathcal{C}}\ell _{1,3}$ is written as 
\begin{eqnarray}
\Upsilon &=&c+c^{0}\mathbf{e}_{0}+c^{1}\mathbf{e}_{1}+c^{2}\mathbf{e}%
_{2}+c^{3}\mathbf{e}_{3}+c^{01}\mathbf{e}_{01}+c^{02}\mathbf{e}_{02}+c^{03}%
\mathbf{e}_{03}+c^{12}\mathbf{e}_{12}+c^{13}\mathbf{e}_{13}  \notag \\
&&+c^{23}\mathbf{e}_{23}+c^{012}\mathbf{e}_{012}+c^{013}\mathbf{e}%
_{013}+c^{023}\mathbf{e}_{023}+c^{123}\mathbf{e}_{123}+c^{0123}\mathbf{e}%
_{0123}.  \notag
\end{eqnarray}%
where $i\mathbf{e}_{0123}=\mathbf{e}_{5}$, and $\mathbf{e}_{\mu }\mathbf{e}%
_{5}=-\mathbf{e}_{5}\mathbf{e}_{\mu }$. From the isomorphism ${\mathcal{C}}%
\ell _{1,3}\simeq {\mathcal{M}}(2,\mathbb{H})$, in order to obtain a
representation of ${\mathcal{C}}\ell _{1,3}$ the primitive idempotent $f=%
\frac{1}{2}(1+\mathbf{e}_{0})$ is used. The left minimal ideal of ${\mathcal{%
C}}\ell _{1,3}$ is written as $I_{1,3}={\mathcal{C}}\ell _{1,3}f$, and an
arbitrary element of $I_{1,3}$ is given by 
\begin{equation}
I_{1,3}\ni \Xi =(a^{1}+a^{2}\mathbf{e}_{23}+a^{3}\mathbf{e}_{31}+a^{4}%
\mathbf{e}_{12})f+(a^{5}+a^{6}\mathbf{e}_{23}+a^{7}\mathbf{e}_{31}+a^{8}%
\mathbf{e}_{12})\mathbf{e}_{5}f,  \notag
\end{equation}%
\noindent where \cite{meudirac} 
\begin{eqnarray}
a^{1} &=&c+c^{0},\qquad a^{2}=c^{23}+c^{023},\qquad
a^{3}=-c^{13}-c^{013},\qquad a^{4}=c^{12}+c^{012},  \notag \\
a^{5} &=&-c^{123}+c^{0123},\qquad a^{6}=c^{1}-c^{01},\qquad
a^{7}=c^{2}-c^{02},\qquad a^{8}=c^{3}-c^{03}.  \notag
\end{eqnarray}%
\noindent Denoting $\mathfrak{i}=\mathbf{e}_{23},\;\mathfrak{j}=\mathbf{e}%
_{31},$ and $\mathfrak{k}=\mathbf{e}_{12}$ it is immediate to see that the $%
\{\mathfrak{i},\mathfrak{j},\mathfrak{k}\}$ satisfies the quaternion algebra
--- under the Clifford product --- and 
\begin{equation}
{\mathcal{C}}\ell _{1,3}f=I_{1,3}\ni \Xi =(a^{1}+a^{2}\mathfrak{i}+a^{3}%
\mathfrak{j}+a^{4}\mathfrak{k})f+(a^{5}+a^{6}\mathfrak{i}+a^{7}\mathfrak{j}%
+a^{8}\mathfrak{k})e_{5}f.  \notag
\end{equation}%
The set $\{1,e_{5}\}f$ is a basis for the ideal $I_{1,3}$, being possible to
write $\mathbf{e}_{\mu }=f\mathbf{e}_{\mu }f+f\mathbf{e}_{\mu }\mathbf{e}%
_{5}f-f\mathbf{e}_{5}\mathbf{e}_{\mu }f-f\mathbf{e}_{5}\mathbf{e}_{\mu }%
\mathbf{e}_{5}f.$ The matrix representation of the orthonormal basis $e_{\mu
}$ is then obtained: 
\begin{equation}
\mathbf{e}_{0}=%
\begin{pmatrix}
1 & 0 \\ 
0 & -1%
\end{pmatrix}%
,\;\;\mathbf{e}_{1}=%
\begin{pmatrix}
0 & \mathfrak{i} \\ 
\mathfrak{i} & 0%
\end{pmatrix}%
,\;\;\mathbf{e}_{2}=%
\begin{pmatrix}
0 & \mathfrak{j} \\ 
\mathfrak{j} & 0%
\end{pmatrix}%
,\;\;\mathbf{e}_{3}=%
\begin{pmatrix}
0 & \mathfrak{k} \\ 
\mathfrak{k} & 0%
\end{pmatrix}%
\end{equation}%
\noindent and the idempotents $f=%
\begin{pmatrix}
1 & 0 \\ 
0 & 0%
\end{pmatrix}%
,\;\,\mathbf{e}_{5}f=%
\begin{pmatrix}
0 & 0 \\ 
1 & 0%
\end{pmatrix}%
$ can be obtained.

Using these representations, it is possible to write $\Upsilon\in{\mathcal{C}%
}\ell_{1,3}$ as 
\begin{eqnarray}  \label{quat}
\mathbf{\Upsilon} &=& \left(%
\begin{array}{cc}
\begin{array}{c}
(c + c^0) + (c^{23} + c^{023})\mathfrak{i} \\ 
+(-c^{13} - c^{013})\mathfrak{j} + (c^{12} + c^{012})\mathfrak{k} \\ 
\quad\quad\quad\quad\quad\quad\quad \\ 
(-c^{123} + c^{0123}) + (c^1 - c^{01})\mathfrak{i} \\ 
+(c^2 - c^{02})\mathfrak{j} + (c^3 - c^{03})\mathfrak{k}%
\end{array}
\begin{array}{c}
(-c^{123} - c^{0123}) + (c^1 + c^{01})\mathfrak{i} + \\ 
(c^2 + c^{02})\mathfrak{j} + (c^3 + c^{03})\mathfrak{k} \\ 
\quad\quad\quad\quad\quad\quad\quad \\ 
(c - c^0) + (c^{23} - c^{023})\mathfrak{i} + \\ 
(-c^{13} + c^{013})\mathfrak{j} + (c^{12} - c^{012})\mathfrak{k}%
\end{array}
& 
\end{array}%
\right)  \notag \\
&=& \left(%
\begin{array}{cc}
q_1 & q_2 \\ 
q_3 & q_4%
\end{array}%
\right).
\end{eqnarray}

In terms of the reversion in ${\mathcal{C}}\ell_{1,3}$ the matrix
representation of $\mathbf{\Upsilon}$ is given by 
\begin{equation}
{\tilde{\mathbf{\Upsilon}}} = \left(%
\begin{array}{cc}
{\bar q}_1 & -{\bar q}_3 \\ 
-{\bar q}_2 & {\bar q}_4%
\end{array}%
\right),
\end{equation}
\noindent where ${\bar q}$ denotes the quaternionic conjugation.

The problem of representing spinor fields by completely skew-symmetric
tensor fields (differential forms) comes back to Ivanenko, Landau and Fock
in 1928, and was considered several times \cite%
{jayme,jayme1,op,lou1,lou2,lase}. An element $\Psi \in \mathcal{C}\ell
_{1,3}^{+}$ --- which corresponds to an operator spinor --- can be written
as 
\begin{equation}
\mathcal{C}\ell _{1,3}^{+}\ni {\Psi }%
=c+c^{01}e_{01}+c^{02}e_{02}+c^{03}e_{03}+c^{12}e_{12}+c^{13}e_{13}+c^{23}e_{23}+c^{0123}e_{0123}.
\label{400}
\end{equation}%
\noindent which in the light of the quaternionic representation in Eq.(\ref%
{quat}) is given by 
\begin{equation}
\left( 
\begin{array}{cc}
q_{1} & -q_{2} \\ 
q_{2} & q_{1}%
\end{array}%
\right) =\left( 
\begin{array}{cc}
c+c^{23}\mathfrak{i}-c^{13}\mathfrak{j}+c^{12}\mathfrak{k} & \quad
c^{0123}-c^{01}\mathfrak{i}-c^{02}\mathfrak{j}-c^{03}\mathfrak{k} \\ 
-c^{0123}+c^{01}\mathfrak{i}+c^{02}\mathfrak{j}+c^{03}\mathfrak{k} & \quad
c+c^{23}\mathfrak{i}+-c^{13}\mathfrak{j}+c^{12}\mathfrak{k}%
\end{array}%
\right) .  \notag
\end{equation}%
Now, considering the isomorphism ${\mathcal{C}}\ell _{1,3}^{+}\simeq {%
\mathcal{C}}\ell _{3,0}\simeq {\mathcal{C}}\ell _{1,3}\frac{1}{2}%
(1+e_{0})\simeq \mathbb{C}^{4}\simeq \mathbb{H}^{2}$, it explicits the
equivalence among the classical, the operatorial, and the algebraic
definitions of a spinor \cite%
{benn,moro,rod,cru,chev,op,hes,hes1,cartan,riesz}. In this sense, the spinor
space $\mathbb{H}^{2}$ carries the $D^{(1/2,0)}\oplus D^{(0,1/2)}$ or $%
D^{(1/2,0)}$, or $D^{(0,1/2)}$ representations of SL(2,$\mathbb{C)}$, it is
isomorphic to the minimal left ideal ${\mathcal{C}}\ell _{1,3}\frac{1}{2}%
(1+e_{0})$ --- corresponding to the algebraic spinor --- and also isomorphic
to the even subalgebra ${\mathcal{C}}\ell _{1,3}^{+}$ --- corresponding to
the operatorial spinor. It is then possible to write a Dirac spinor field as 
\begin{equation}
\left( 
\begin{array}{cc}
q_{1} & -q_{2} \\ 
q_{2} & q_{1}%
\end{array}%
\right) \frac{1}{2}(1+\mathbf{e}_{0})=\left( 
\begin{array}{c}
c+c^{23}\mathfrak{i}-c^{13}\mathfrak{j}+c^{12}\mathfrak{k} \\ 
c^{0123}-c^{01}\mathfrak{i}-c^{02}\mathfrak{j}-c^{03}\mathfrak{k}%
\end{array}%
\right) \in \mathbb{H}\oplus \mathbb{H}.  \label{hh}
\end{equation}%
\noindent Returning to Eq.(\ref{400}), and using for instance the standard
representation $\Psi $ can be represented by 
\begin{equation*}
\begin{pmatrix}
c - ic^{12} & c^{13} - ic^{23} & -c^{03} + i^{0123} & -c^{01} + i c^{02} \\ 
-c^{13} - ic^{23} & c + ic^{12} & -c^{01} - i c^{02} & c^{03} + ic^{0123} \\ 
-c^{03} + ic^{0123} & -c^{01} + i c^{02} & c - ic^{12} & c^{13} - ic^{23} \\ 
-c^{01} - i c^{02} & c^{03} + ic^{0123} & -c^{13} - ic^{23} & c + ic^{12}%
\end{pmatrix}%
:=%
\begin{pmatrix}
\phi_1 & -\phi_2^* & \phi_3 & \phi_4^* \\ 
\phi_2 & \phi_1^* & \phi_4 & -\phi_3^* \\ 
\phi_3 & \phi_4^* & \phi_1 & -\phi_2^* \\ 
\phi_4 & -\phi_3^* & \phi_2 & \phi_1^*%
\end{pmatrix}%
.
\end{equation*}%
The Dirac spinor $\psi $ is an element of the minimal left ideal $(\mathbb{C}%
\otimes \mathcal{C}\ell _{1,3})f$, where\footnote{%
We choose to express $f=\frac{1}{4}(1+\gamma _{0})(1+i\gamma _{12})$ using
the Dirac representation. It could be chosen the idempotent $f=\frac{1}{4}%
(1+\gamma _{5})(1+i\gamma _{12})$ associated with the Weyl representation.} $%
f=\frac{1}{4}(1+\gamma _{0})(1+i\gamma _{12})$. We choose take the Dirac
standard representation that sends the basis vectors $\mathbf{e}_{\mu }$ to $%
\gamma _{\mu }\in $ End($\mathbb{C}^{4}$). Then, 
\begin{equation}
\psi =\Phi \frac{1}{2}(1+i\gamma _{12})\in (\mathbb{C}\otimes \mathcal{C}%
\ell _{1,3})f,
\end{equation}%
\noindent where $\Phi =\Phi \frac{1}{2}(1+\gamma _{0})\in \mathcal{C}\ell
_{1,3}(1+\gamma _{0})$ is two times the real part of $\psi $. Using the
matrix representation it follows that 
\begin{equation}
(\mathbb{C}\otimes \mathcal{C}\ell _{1,3})f\ni \psi \simeq \mathbb{C}\otimes 
\begin{pmatrix}
\phi_1 & 0 & 0 & 0 \\ 
\phi_2 & 0 & 0 & 0 \\ 
\phi_3 & 0 & 0 & 0 \\ 
\phi_4 & 0 & 0 & 0%
\end{pmatrix}%
\simeq \mathbb{C}\otimes 
\begin{pmatrix}
\phi_1 \\ 
\phi_2 \\ 
\phi_3 \\ 
\phi_4%
\end{pmatrix}%
=%
\begin{pmatrix}
\psi_1 \\ 
\psi_2 \\ 
\psi_3 \\ 
\psi_4%
\end{pmatrix}%
\in \mathbb{C}^{4},  \label{dire}
\end{equation}%
\noindent where it can be seen the direct correspondence between $\psi $ and
the classical Dirac spinor.

\section{Mapping Dirac to ELKO spinor fields and the instanton Hopf fibration%
}

\label{elkoo} The suitable mathematical structure to describe the instanton
is a principal bundle with base manifold $S^4$ and associated structural
group SU(2). In \cite{jayme} a formalism
similar to the magnetic monopole  was exhibited and constructed, exploring the relationship between spinor
fields and the bilinear covariants. Spinor fields indirectly describe
fermionic fields, since the observables are their associated bilinear
covariants. In \cite{qtmosna} a tomographic scheme --- based on spacetime
symmetries --- was presented for the reconstruction of the internal degrees
of freedom of a Dirac spinor, together with the possibility of the
tomographic group be taken as SU(2). In addition, in \cite{cra} the spinor
field was reconstructed from the bilinear covariants.

As argued in \cite{jayme}, using the inversion theorem for Euclidean
signature, it is possible to formulate those constructions for the case of
magnetic monopoles and instantons, indicating the generalizations of the
Balachandran's construction to the case of instantons \cite{bala}. Also, the
inversion theorem for Minkowski spacetime appeared for the first time in the
paper \cite{balawal}. On the other hand, in a previous paper \cite{osmano}
we investigate and provide the necessary and sufficient conditions to map
Dirac spinor fields (DSFs) to ELKO, in order to naturally extend the
Standard Model to spinor fields possessing mass dimension one. Let us make a
briefly review of which are the conditions a Dirac spinor field must obey to
be led to an ELKO. In \cite{osmano} there has been proved that not all DSFs
can be led to ELKO, but only a subset of the three classes --- under
Lounesto classification --- of DSFs restricted to some conditions.
Explicitly, by taking a DSF 
\begin{equation}
\psi(\mathbf{p}) = 
\begin{pmatrix}
\phi_{R}(\mathbf{p}) \\ 
\phi_{L}(\mathbf{p})%
\end{pmatrix}
=%
\begin{pmatrix}
\epsilon \sigma_{2}\phi^{*}_{L}(\mathbf{p}) \\ 
\phi_{L}(\mathbf{p})%
\end{pmatrix}
,  \label{comeco}
\end{equation}
and taking into account that $\phi_{R}(\mathbf{p})=\chi \phi_{L}(\mathbf{p})$%
, where $\chi = \frac{E + {\mathbf{\sigma}}\cdot {\mathbf{p}}}{m}$ and $%
\kappa\psi = \psi^*$, and denoting the 4-component DSF by $\psi =
(\psi_1,\psi_2,\psi_3,\psi_4)^T$ ($\psi_r \in \mathbb{C}, r=1,\ldots,4$), we
have the simultaneous conditions a DSF must obey in order for it to be led
to an ELKO \cite{osmano}: 
\begin{eqnarray}
0&=& \mathbb{R}\mathrm{e}(\psi_1^*\psi_3)=\mathbb{R}\mathrm{e}%
(\psi_2^*\psi_4)  \notag \\
0&=&\mathbb{R}\mathrm{e} (\psi_2^*\psi_3)+ \mathbb{R}\mathrm{e}
(\psi_1^*\psi_4)  \notag \\
0&=& \mathrm{Im}(\psi_1^*\psi_4)-\mathrm{Im}(\psi_2^*\psi_3)-2\mathrm{Im}%
(\psi_3^*\psi_4)-2\mathrm{Im}(\psi_1^*\psi_2) .  \label{partes}
\end{eqnarray}
In what follows we obtain the extra necessary and sufficient conditions for
each class of DSFs.

As additional conditions on class-(2) Dirac spinors, we also have: 
\begin{eqnarray}
\mathbb{R}\mathrm{e}(\psi_1^*\psi_4)+\mathrm{Im}(\psi_2^*\psi_3) &=&0 .
\label{ad2}
\end{eqnarray}

For the class-(3) of spinor fields, the additional condition was obtained in 
\cite{osmano}: 
\begin{eqnarray}
\mathrm{Im}(\psi_1^*\psi_4)-\mathrm{Im}(\psi_2^*\psi_3)-2\mathrm{Im}%
(\psi_1^*\psi_2) &=&0 .  \label{ad3}
\end{eqnarray}

Class-(1) DSFs must obey all the conditions given by Eqs.(\ref{partes}), (%
\ref{ad2}), and (\ref{ad3}). Note that if one relaxes the condition given by
Eq.(\ref{ad2}) or Eq.(\ref{ad3}), DSFs of types-(3) and -(2) are
respectively obtained.

Using the decomposition $\psi_j=\psi_{ja}+i\psi_{jb}$ (where $\psi_{ja}$ = $%
\mathbb{R}$e($\psi_j$) and $\psi_{jb}$ = Im($\psi_j$)) it follows that $%
\mathbb{R}\mathrm{e}(\psi_i^*\psi_j)=\psi_{ia}\psi_{ja}+\psi_{ib}\psi_{jb}$
and $\mathrm{Im}(\psi_i^*\psi_j)=\psi_{ia}\psi_{jb}-\psi_{ib}\psi_{ja}$ for $%
i,j=1,\ldots,4$. So, in components, the conditions in common for all types
of DSFs are 
\begin{eqnarray}
\psi_{1a}\psi_{3a}+\psi_{1b}\psi_{3b}&=&0 ,  \label{c1} \\
\psi_{2a}\psi_{4a}+\psi_{2b}\psi_{4b}&=&0 ,  \label{c2}
\end{eqnarray}
and the additional conditions for each case are summarized in Table I below.

\begin{center}
\begin{table}[!h]
\begin{tabular}{|c|c|}
\hline
\textbf{Class} & \textbf{Additional conditions} \\ \hline\hline
(1) & $\psi_{2a}(\psi_{3a}-\psi_{3b})+\psi_{2b}(\psi_{3a}+\psi_{3b}) = 0 =
\psi_{3a}\psi_{4b}-\psi_{3b}\psi_{4a}$ \\ \hline
(2) & $\psi_{3a}\psi_{4b}-\psi_{3b}\psi_{4a} = 0 =
\psi_{2a}\psi_{3a}+\psi_{2b}\psi_{3b}+\psi_{1a}\psi_{4a}+\psi_{1b}\psi_{4b}$
\\ \hline
(3) & $\psi_{2a}(\psi_{3a}-\psi_{3b})+\psi_{2b}(\psi_{3a}+\psi_{3b})=0$ and
\\ 
{} & $(\psi_{1a}\psi_{4b}-\psi_{1b}\psi_{4a})-(\psi_{2a}\psi_{3b}-\psi_{2b}%
\psi_{3a})$ \\ 
& $-2(\psi_{3a}\psi_{4b}-\psi_{3b}\psi_{4a})-$ $2(\psi_{1a}\psi_{2b}-%
\psi_{1b}\psi_{2a})=0$ \\ \hline
\end{tabular}%
\medbreak
\caption{Additional conditions, in components, for class (1), (2) and (3)
Dirac spinor fields.}
\end{table}
\end{center}

The explicit mappings obtained above present the same form of the instanton
Hopf fibration map $S^3\ldots S^7 \rightarrow S^4$ mapping obtained in \cite%
{jayme}, and could be interpreted as the geometric meaning of the mass
dimension-transmuting operator obtained in \cite{osmano}, where we obtained
mapping between ELKO and Dirac spinor fields. As the latter possess mass
dimension 3/2, the former presents mass dimension 1. Some results involving
the instanton Hopf fibration can also be seen in this context, e.g, in \cite%
{jal}. It could explain why ELKO spinor fields satisfy a Klein-Gordon
equation, instead of the Dirac equation \cite{allu,alu2,alu4,alu3,boe1,gau}.

Indeed, the monopole construction was based \cite{jayme} on the Hopf
fibration $S^1\ldots S^3\rightarrow S^2$, where $S^1$ is homeomorphic to the
Lie gauge group U(1) of the electromagnetism. Using a similar construction 
\cite{jayme}, the instanton is related to a principal bundle with structure
Lie group SU(2), which is homeomorphic to the 3-sphere $S^3$. The instanton
was described in \cite{jayme} using the the Hopf fibration $S^3\ldots
S^7\rightarrow S^4$ by means of the bilinear covariants associated with the
Dirac spinor fields, under Lounesto spinor field classification.

Types-(1) and -(2) Dirac spinor fields can be regarded as satisfying $%
\sigma=1$, which is exactly $S^7$, when the Dirac spinor field is
classically described by an element of $\mathbb{C}^4\simeq\mathbb{H}^2$ (here 
these spaces are isomorphic as vector spaces).
Now, the Fierz identities described in Eq.(\ref{fi}) give immediately ---
from the equation $\sigma=1$ --- the expression $\mathbf{J}^2 + \omega^2 =1$%
, which is $S^4$.

The mapping in Eq.(\ref{hh}) induces the possibility to interpret the
coordinate 8-tuple ($c, c^{23},-c^{13},c^{12},c^{0123}, - c^{01},-c^{02},-
c^{03}$) as local coordinates in $S^7$, and then $S^7$ is the (compact)
space described by an unitary Dirac spinor. The Fierz identities imply that $%
\mathbf{J}^2 + \omega^2 =1$, which is topologically an $S^4$ with local
coordinates $(J_0,J_1,J_2,J_3,\omega)$.

Using the definition of the bilinear covariants in Eq.(\ref{fierz}) and the
quaternionic representation of the Dirac spinor in Eq.(\ref{hh}), it is
possible to write \cite{jayme} 
\begin{eqnarray}
\sigma &=& \|q_1\|^2 + \|q_2^2\|,\qquad \omega = 2\,\mathrm{Re}(q_1^\ast
q_2),\qquad J_0 = \|q_1\|^2 - \|q_2^2\|  \notag \\
J_1 &=& 2\,\mathrm{Re}(q_1^\ast\, \mathfrak{i}\, q_2),\qquad J_2 = 2\,%
\mathrm{Re}(q_1^\ast\, \mathfrak{j}\, q_2),\qquad J_3 = 2\,\mathrm{Re}%
(q_1^\ast\, \mathfrak{k}\, q_2),  \notag
\end{eqnarray}
\noindent which in the representation given by Eq.(\ref{dire}) is given by 
\cite{jayme} 
\begin{eqnarray}
\sigma &=& \|\psi_1\|^2 + \|\psi_2\|^2 + \|\psi_3\|^2 + \|\psi_4\|^2 = 1 
\notag \\
J_0 &=& \|\psi_1\|^2 + \|\psi_2\|^2 - \|\psi_3\|^2 - \|\psi_4\|^2 = 1  \notag
\\
J_1 &=& 2 \mathrm{Im}(\psi_1\psi_4^\ast) + 2 \mathrm{Im}(\psi_2\psi_3^\ast) 
\notag \\
J_2 &=& 2 \mathrm{Re}(\psi_2\psi_3^\ast) - 2 \mathrm{Re}(\psi_1\psi_4^\ast) 
\notag \\
J_3 &=& 2 \mathrm{Im}(\psi_3\psi_1^\ast) + 2 \mathrm{Im}(\psi_2\psi_4^\ast) 
\notag \\
\omega &=& 2 \mathrm{Re}(\psi_1\psi_3^\ast) + 2 \mathrm{Re}%
(\psi_2\psi_4^\ast) .  \label{s44}
\end{eqnarray}%
\noindent Although these expressions are not the same as Eqs.(\ref{partes}, %
\ref{ad2}, \ref{ad3}) we might argue whether there is a corresponding
application $M\in$ End($\mathbb{C}^4$) leading Dirac to ELKO spinor fields 
 that indeed corresponds to the expressions above, in the light
of the procedure in \cite{osmano}. In the paper \cite{osmano} there is an
explicit algorithm that constructs such an application, using
straightforward assumptions. Even if we could keep the same application $%
M\in $ End($\mathbb{C}^4$) as obtained in \cite{osmano} and change the form
of the Dirac spinor field, or take the same spinor field in Eq.(\ref{dire})
and construct another application $M^\prime\in$ End($\mathbb{C}^4$) ---
using an analogous procedure as explicitly exhibited in \cite{osmano} --- in
such a way that the instanton Hopf fibration conditions in Eqs.(\ref{s44})
and the Dirac to ELKO mapping in Eqs.(\ref{partes}, \ref{ad2}, \ref{ad3}) be
similar, we remember that in Eqs.(\ref{s44}) the terms $J_0, J_1, J_2, J_3$
cannot simultaneously equal zero, because $\mathbf{J} \neq 0$. Formally, the
instanton cannot be described by an ELKO spinor field. This statement
mathematically explain the well known physical interpretation that while the
instanton is a localized topological object, ELKO is a non-local extended
one \cite{allu}.

\section{Type-(4) (flag-dipole) spinor fields}

It has been argued that the flag-dipole spinor fields (type-(4) under
Lounesto spinor field classification) are related to the quark confinement,
although they are not appropriate to describe fermions, since they do not
constitute a real vector space \cite{lou1}. As ELKO spinor fields are prime
candidates to describe dark matter, and the flag-dipole spinor fields can
shed some new light on the quark confinement investigations, we want to
point out some algebraic and geometric considerations concerning the
type-(4) spinor fields.

The Weyl and Majorana spinor fields can be written in terms of operator
spinor fields as 
\begin{equation*}
\Psi \frac{1}{2}(1+\gamma _{0}{\mathbf{u}}),\qquad \Psi \in \sec \mathcal{C}%
^{+}\ell (M,\eta ),
\end{equation*}%
where $\mathcal{C}^{+}\ell (M,\eta )$ denotes the spacetime Clifford bundle,
in which the typical fiber is $\mathcal{C}^{+}\ell _{1,3}$ where $\mathbf{u}%
= $ $\pm \gamma _{2}$ for Weyl spinor fields and $\mathbf{u}=$ $\pm \gamma
_{1} $ for Majorana spinor fields.

More generally, ELKO spinor fields can also be written in the same form, as 
\begin{equation*}
\Psi\frac{1}{2}(1+\gamma_0{\mathbf{u}}),\qquad\Psi\in \sec \mathcal{C}%
^+\ell(M,\eta)
\end{equation*}
where \textbf{u} propitiates a mixture of Weyl and Majorana spinor fields (%
\textbf{u} = $\gamma_1\cos\alpha + \gamma_2\sin\alpha$). This mixture can be
written as $\mathbf{u}=\gamma_1\cos\phi + \mathbf{i}\gamma_3\sin\phi$, where 
\textbf{i} = $-\gamma_2\gamma_3$.

In addition, following Doran's conjecture \cite{lou1}, all the flag-dipole
--- type-(4) spinor fields under Lounesto spinor field classification ---
can be written in a similar form as 
\begin{eqnarray}
\Psi\frac{1}{2}(1+\gamma_0{\mathbf{u}}),\qquad\Psi\in \mathcal{C}%
^+\ell(M,\eta),\quad \mathbf{u}\in\mathbb{R}^3,\quad \mathbf{u}^2=-1.  \notag
\end{eqnarray}%
\noindent More precisely, it is assumed that $\mathbf{u}$ is a spatial unit
vector (\textbf{u}$\cdot\gamma_0=-1$), and \textbf{u} is neither a multiple
of $\gamma_3$ nor a multiple of $\gamma_1\gamma_2$.

Now, by introducing the complex multivector field as in \cite{lou1,lou2} $%
Z\in\sec\mathbb{C}\ell(M,\eta)$ (where $\mathbb{C}\ell(M,\eta)$ denotes the
complexified spacetime Clifford bundle, in which the typical fiber is $%
\mathbb{C\otimes \mathcal{C}\ell}_{1,3}\simeq \mathcal{C}\ell_{4,1}$ \cite%
{moro}) and the corresponding complex multivector\textit{\ operator }%
(represented by the same letter): 
\begin{equation}
Z=\sigma+\mathbf{J}+i\mathbf{S}+i\mathbf{K}\gamma_{0123}+\omega\gamma_{0123}.
\label{boom}
\end{equation}
When the multivector operators $\sigma,\omega,\mathbf{J},\mathbf{S},\mathbf{K%
}$ satisfy the Fierz identities, then the complex multivector operator $Z$
is denominated a \emph{Fierz aggregate}, and, when $\gamma
_{0}Z^{\dagger}\gamma_{0}=Z$, which means that $Z$ is a Dirac self-adjoint
aggregate\footnote{%
It is equivalent to say that $\omega,\sigma,\mathbf{J},\mathbf{K},\mathbf{S}$
are real multivector fields.}, $Z$ is called a \emph{boomerang}.

A spinor field such that \emph{not both} $\omega$ and $\sigma$ are null is
said to be regular. When $\omega=0=\sigma$, a spinor field is said to be 
\textit{singular}. In this case the Fierz identities are in general replaced
by the more general conditions \cite{cra} (which obviously also hold for $%
\omega,\sigma\neq0$). These conditions are: 
\begin{align}
Z^{2} & =4\sigma Z,\qquad Z\gamma_{\mu}Z=4J_{\mu}Z,\qquad Zi\gamma_{\mu\nu
}Z=4S_{\mu\nu}Z,  \notag \\
Zi\gamma_{0123}\gamma_{\mu}Z & =4K_{\mu}Z,\qquad Z\gamma_{0123}Z=-4\omega Z.
\end{align}

Now, any spinor field (regular or singular) can be reconstructed from its
bilinear covariants as follows. Take an arbitrary spinor field $\xi$
satisfying $\xi^{\dagger}\gamma_{0}\psi\neq0.$ Then the spinor field $\psi$
and the multivector field $Z\xi$, differ only by a phase. Indeed, it can be
written as 
\begin{equation}
\psi=\frac{1}{4N}e^{-i\alpha}Z\xi,  \notag
\end{equation}
\noindent where $N=\frac{1}{2}\sqrt{\xi^{\dagger}\gamma_{0}Z\xi}$ and $%
e^{-i\alpha}=\frac{1}{N}\xi^{\dagger}\gamma_{0}\psi$. For more details see,
e.g., \cite{cra, qtmosna}.

For the specific case of type-(4) (flag-dipole) spinor fields, the boomerang
can be written as 
\begin{eqnarray}  \label{boom1}
Z=\mathbf{J} + i\mathbf{J}s - ih\gamma_{0123}\mathbf{J},\qquad \mathbf{J}^2=0
\end{eqnarray}%
\noindent where $s$ denotes a spacelike vector orthogonal to \textbf{J},
meaning that $s^2 < 0$ and $\mathbf{J}\lrcorner s = 0$. The bilinear
covariant $\mathbf{S}$ is given by the Clifford product $\mathbf{S} = 
\mathbf{J}s = \mathbf{J}\wedge s$. Exclusively for the type-(4) flag-dipole
spinor fields, the real coefficient satisfies $h\neq 0$. For all other types
of spinor fields, including type-(4), $h$ is constrained to $s$ by the
expression 
\begin{eqnarray}  \label{hs}
h^2 = 1 + s^2 < 1, \quad\text{since $s^2 < 0$}
\end{eqnarray}%
\noindent and is defined as to relate the two bilinear covariants \textbf{K}
and \textbf{J} by \textbf{K} = $h$\textbf{J}.

Using Eq.(\ref{boom1}), it is immediate to verify that $(1 + is -
ih\gamma_{0123})Z=0$ and also that the boomerang $Z$ is a lightlike Clifford
multivector, i.e., $Z^2=0$, since flag-dipole spinor fields in class-(4)
under Lounesto spinor field classification satisfies $\mathbf{J}^2 = 0$ (see
Eq.(\ref{fierz})). Doran's conjecture asserts that the coefficient $h$ is
given by \textbf{u}$\cdot\gamma_3$ \cite{lou1,lou2}. Also, the equation $%
Z^2=0$ implies that $Z=\mathbf{J}(1 + is + ih\gamma_{0123})$.

Furthermore, using the representation of the type-(4) flag-dipole spinor
field $\psi$ as an element of the minimal left ideal $(\mathbb{C}\otimes{%
\mathcal{C}}\ell_{1,3})\frac{1}{2}(1+\gamma_0)\frac{1}{2}(1+i\gamma_1%
\gamma_2)$ it follows that \cite{lou1,lou2} $\frac{1}{2}(1 - is -
ih\gamma_{0123})\Psi = \Psi$, while $\frac{1}{2}(1 + is +
ih\gamma_{0123})\Psi = 0$.

Now, by means of the isomorphism ${\mathcal{C}}\ell_{1,3}^+\simeq{\mathcal{C}%
}\ell_{3,0}\simeq {\mathcal{C}}\ell_{1,3}\frac{1}{2}(1+\gamma_0)\simeq%
\mathbb{C}^4$ Lounesto defined a projector $\Sigma_\pm\in$ End(${\mathcal{C}}%
\ell_{1,3}$) by the expression 
\begin{eqnarray}  \label{us}
\Sigma_\pm(u)=\frac{1}{2}(u\pm (s\mp \gamma_{0123}\mathbf{J}\mathbf{K}^{-1}
u\gamma_{0123})),
\end{eqnarray}%
\noindent in such a way that this definition keep unaltered an ideal
(algebraic) spinor in the minimal left ideal $\psi\in(\mathbb{C}\otimes{%
\mathcal{C}}\ell_{1,3})f$. As for ideal spinor fields $\psi$ the equation $%
\psi\gamma_{0123} = \psi\gamma_2\gamma_1$, for the case where $\psi$ is a
type-(4) flag-dipole spinor $\Psi$, the relation $\frac{1}{2}(1 + is +
ih\gamma_{0123})\Psi = 0$ holds, as we have just seen. In this case, $%
\mathbf{K} = h\mathbf{J}$, and the projector $\Sigma_\pm\in$ End(${\mathcal{C%
}}\ell_{1,3}$) acts on $\Psi$ as 
\begin{eqnarray}  \label{us1}
\Sigma_\pm(\Psi)=\frac{1}{2}(\Psi\pm
(s+h\gamma_{0123})\Psi\gamma_{1}\gamma_2).
\end{eqnarray}%
\noindent

ELKO spinors $\lambda(\mathbf{p})$ are obtained as a particular case where $%
h=0$. Indeed, as type-(4) spinor fields are defined by the relations assumed
by their bilinear covariants $\omega=0=\sigma$, \textbf{K} $\neq 0$, \textbf{%
J} $\neq 0$. As $\mathbf{K} = h\mathbf{J}$, when we put $h=0$, the bilinear
covariants assume the expressions $\omega=0=\sigma$, \textbf{K} $= 0$, 
\textbf{J} $\neq 0$, which are precisely the bilinear covariants associated
with ELKO (type-(5)) spinor fields. Then, ELKO spinor fields can be thought
as being limiting cases of type-(4) spinor fields: 
\begin{eqnarray}
\lambda(\mathbf{p}) = \Sigma_\pm(\Psi)=\frac{1}{2}(\Psi\pm
s\Psi\gamma_{1}\gamma_2).  \label{45}
\end{eqnarray}%
\noindent Clearly, all the six classes under Lounesto spinor field
classification are disjoint classes, and in particular, type-(4) flag-dipole
spinor fields and type-(5) flagpole (ELKO) spinor fields are disjoint. The
limit $h\rightarrow 0$ changes the class (4) into class (5).

Also, using Eq.(\ref{hs}), it is also possible to turn type-(4) flag-dipole
spinor fields into type-(6) Weyl spinor fields, in the limit $s\rightarrow 0$
--- implying that $h=\pm 1$, and Weyl spinor fields can be alternatively
written as \cite{lou1,lou2} 
\begin{eqnarray}  \label{us2}
\Sigma_\pm(\Psi)=\frac{1}{2}(\Psi\pm \gamma_{0123}\Psi\gamma_{1}\gamma_2).
\end{eqnarray}%
\noindent In this sense, endomorphisms of ${\mathcal{C}}\ell_{1,3}$ where
exhibited in \cite{lou1,lou2}, which make possible to change spinor fields
classes under Lounesto spinor field classification. Weyl and ELKO spinor
fields are obtained from an arbitrary type-(4) flag-dipole spinor field,
respectively corresponding to type-(6) and type-(5) spinor fields.

\subsection*{Acknowledgment}

The authors are very grateful to Prof. Dharamvir Ahluwalia-Khalilova for
important comments about this paper, and also to the AACA referee for the
valuable and relevant comments and suggestions that improved this
manuscript. J. M. Hoff da Silva thanks to CAPES-Brazil for financial
support. 

\section*{Appendix A: Operator Spinors}

Recall that given the $\mathbb{Z}_2$-graded Clifford algebra $\mathcal{C}%
\ell_{1,3}$, we can use the even subalgebra ${\mathcal{C}}\ell_{1,3}^+\simeq{%
\mathcal{C}}\ell_{3,0}$ as the representation space for $\mathcal{C}%
\ell_{p,q}$. Define a representation $\rho: \mathcal{C}\ell_{1,3}
\rightarrow \mathrm{End}(\mathcal{C}\ell_{1,3}^+)$, the so-called
irreducible graded representation (IGR).

A multivector $\psi \in \mathcal{C}\ell_{1,3}$ can be split as $\psi =
\psi_+ + \psi_-$, where $\psi_\pm = \frac{1}{2}(\psi \pm \hat{\psi})\in{%
\mathcal{C}}\ell_{1,3}^\pm. $ Consider now $\rho = \rho_+ + \rho_-$ and $%
\rho (\psi) = \rho_+ (\psi_+) + \rho_-(\psi_-). $ For $\psi_- \in \mathcal{C}%
\ell_{p,q}^-$ it follows that $a_-\phi \in \mathcal{C}\ell_{1,3}^-$ for $%
\phi \in \mathcal{C}\ell_{1,3}^+$, i.e., $\rho_+(\psi_+)(\phi) =
\psi_+\phi,\quad\forall\phi \in \mathcal{C}\ell_{1,3}^+. $ Now take an odd
element $\varsigma \in\mathcal{C}\ell_{1,3}^-$ and define $%
\rho_-(\psi_-)(\phi) = \psi_-\phi\varsigma,\quad\forall\phi \in \mathcal{C}%
\ell_{1,3}^+$. If $\varsigma$ is chosen in such a way that $\varsigma^2 = 1,$
where $\varsigma \in \mathcal{C}\ell_{1,3}^-, $ the definition of IGR does
depend on the existence of an odd element such that $\varsigma^2 = 1$. In
the particular cases $\mathcal{C}\ell_{0,1} \simeq \mathbb{C}$ and $\mathcal{%
C}\ell_{0,2} \simeq \mathbb{H}$, such an element does not exist. In order to
show that $\rho$ is irreducible, suppose that there exists an element $%
\varpi_1 \in \mathcal{C}\ell_{p,q}^+$ such that $(\varsigma_1)^2 = 1$ and $%
\varpi\varsigma = \varsigma\varpi$. We write $\mathcal{C}\ell_{1,3}^+ = {}_+%
\mathcal{C}\ell_{1,3}^+ \oplus {}_-\mathcal{C}\ell_{1,3}^+,$ where ${}_\pm%
\mathcal{C}\ell_{1,3}^+ = \mathcal{C}\ell_{1,3}^+ \frac{1}{2}(1 \pm
\varpi_1) $, and, for $\phi_\pm \in {}_\pm\mathcal{C}\ell_{1,3}^+$, it
follows that $\phi_\pm\varpi_1 = \pm\phi_\pm. $ Each one of the spaces $%
{}_\pm\mathcal{C}\ell_{1,3}^+$ is invariant under $\rho$, as can be
immediately seen by the relations $(\varsigma_1)^2 = 1$ and $\varpi\varsigma
= \varsigma\varpi$, and in addition these subespaces are subalgebras of $%
\mathcal{C}\ell_{1,3}^+$.

If there exists another even element $\varpi_2$ such that $(\varpi_2)^2 = 1$%
, then $\varpi_2\varpi_1 = \varpi_1\varpi_2$ and $\varpi_2\varsigma =
\varsigma\varpi_2$. It follows that the subspaces ${}_\pm\mathcal{C}%
\ell_{1,3}^+$ do not carry an irreducible representation. It is defined 
\begin{eqnarray}
{}_\pm{}_\pm\mathcal{C}\ell_{1,3}^+ = {}_\pm\mathcal{C}\ell_{1,3}^+ \frac{1}{%
2}(1 \pm \varpi_1)\frac{1}{2}(1 \pm \varpi_2),
\end{eqnarray}
each one is invariant under $\rho$, i.e., $\rho({}_\pm{}_\pm\mathcal{C}%
\ell_{1,3}^+)\hookrightarrow {}_\pm{}_\pm\mathcal{C}\ell_{p,q}^+$. It is
possible to continue this construction in $\mathcal{C}\ell_{p,q}$ when there
is another even element $\varpi_3$ such that $(\varpi_3)^2 = 1$, $%
\varpi_3\varpi_1 = \varpi_1\varpi_3$, $\varpi_3\varpi_2 = \varpi_2\varpi_3$
and $\varpi_3\varsigma = \varsigma\varpi_3$.

When there is not even elements satisfying these conditions anymore, an
irreducible representation is obtained. The space that carries such
representations is called spinor algebra, a subalgebra of the even
subalgebra. In some cases it can be the even subalgebra itself. An element
of the IGR of $\mathcal{C}\ell_{1,3}$ is called an operator spinor.


\begin{thebibliography}{99}
\bibitem{allu} D. V. Ahluwalia-Khalilova and D. Grumiller, \emph{Spin Half
Fermions, with Mass Dimension One: Theory, Phenomenology, and Dark Matter},
J. Cosm. Astrop. Phys. \textit{JCAP} \textbf{07} (2005) 012 [\texttt{%
arXiv:hep-th/0412080v3}].

\bibitem{alu2} D. V. Ahluwalia-Khalilova and D. Grumiller, \emph{Dark
matter: A spin one half fermion field with mass dimension one?}, Phys. Rev. 
\textbf{D 72} (2005) 067701 [\texttt{arXiv:hep-th/0410192v2}].

\bibitem{alu4} D. V. Ahluwalia-Khalilova, \emph{Extended set of Majorana
spinors, a new dispersion relation, and a preferred frame}, [\texttt{%
arXiv:hep-ph/0305336v1}].

\bibitem{ahlu4} D. V. Ahluwalia-Khalilova, \emph{Theory of neutral
particles: Mclennan-Case construct for neutrino, its generalization, and a
new wave equation}, Int. J. Mod. Phys. \textbf{A 11} (1996) 1855-1874 [%
\texttt{arXiv:hep-th/9409134v2}].

\bibitem{lou1} P. Lounesto, \emph{Clifford Algebras, Relativity and Quantum
Mechanics}, in Letelier P and Rodrigues W A, Jr. (eds.), \emph{Gravitation:
the Spacetime Structure}, Proc. of the $8^{\mathrm{th}}$ Latin American
Symposium on Relativity and Gravitation, \'{A}guas de Lind\'{o}ia, Brazil,
25-30 July 1993, World-Scientific, London 1993.

\bibitem{lou2} P. Lounesto, \emph{Clifford Algebras and Spinors}, 2$^{%
\mathrm{nd}}$ ed., pp. 152-173, Cambridge Univ. Press, Cambridge 2002.

\bibitem{meu} R. da Rocha and W. A. Rodrigues, Jr., \emph{Where are ELKO
spinors in Lounesto spinor field classification?}, Mod. Phys. Lett. \textbf{%
A 21} (2006) 65-74 [\texttt{arXiv:math-ph/0506075v3}].

\bibitem{alu3} D. V. Ahluwalia-Khalilova, \emph{Dark matter, and its darkness%
}, Int. J. Mod. Phys. \textbf{D15} (2006) 2267-2278 [\texttt{%
arxiv:hep-th/0603545v3}].

\bibitem{boe1} C. G. Boehmer, {}{The Einstein-Elko system -- Can dark matter
drive inflation?}, \emph{Annalen Phys.} \textbf{16} (2007) 325-341 [\texttt{%
arXiv:gr-qc/0701087v1}]; {}{The Einstein-Cartan-Elko system}, \emph{Annalen
Phys.} \textbf{16} (2007) 38-44 [\texttt{arXiv:gr-qc/0607088v1}]; {}{Dark
spinor inflation -- theory primer and dynamics}, \emph{Phys. Rev.} \textbf{D
77} (2008) 123535 [\texttt{arXiv:0804.0616v1 [astro-ph]}]. 

\bibitem{hor} S. Holst, \emph{Barbero's Hamiltonian derived from a
generalized Hilbert-Palatini action}, Phys. Rev. \textbf{D 53} (1996)
5966-5969 [\texttt{arXiv:gr-qc/9511026v1}].

\bibitem{tn1} J. M. Nester and R. S. Tung, \emph{A Quadratic Spinor
Lagrangian for General Relativity}, Gen. Rel. Grav. \textbf{27} (1995)
115-119 [\texttt{arXiv:gr-qc/9407004v1}].

\bibitem{bars1} I. Bars and S. W. MacDowell, \emph{A spin-3/2 theory of
gravitation}, Gen. Rel. Grav. \textbf{10} (1979) 205-209.

\bibitem{jpe} R. da Rocha and J. G. Pereira, \emph{The quadratic spinor
Lagrangian, axial torsion current, and generalizations}, Int. J. Mod. Phys. 
\textbf{D 16} (2007) 1653-1667 [\texttt{arXiv:gr-qc/0703076v1}].

\bibitem{ro3} R. da Rocha and W. A. Rodrigues, Jr., {}{\ The
Einstein-Hilbert Lagrangian density in a 2-dimensional spacetime is an exact
differential}, \emph{\ Mod. Phys. Lett.} \textbf{A21} (2006) 1519-1527 [%
\texttt{arXiv:hep-th/0512168v7}].

\bibitem{ro4} W. A. Rodrigues, Jr., R. da Rocha, and J. Vaz, Jr., \emph{%
Hidden Consequence of Active Local Lorentz Invariance}, Int. J. Geom. Meth.
Mod. Phys. \textbf{2} (2005) 305-357 [\texttt{arXiv:math-ph/0501064v6}].

\bibitem{ro5} R. da Rocha and W. A. Rodrigues, Jr., {}{\ The Dirac-Hestenes
Equation for Spherical Symmetric Potentials in the Spherical and Cartesian
Gauges}, \emph{\ Int. J. Mod. Phys.} \textbf{A 21} (2006) 4071-4082 [\texttt{%
arXiv:math-ph/0601018v2}].


\bibitem{osmano} R. da Rocha and J. M. Hoff da Silva, \emph{From Dirac
spinor fields to eigenspinoren des ladungskonjugationsoperators}, {J. Math.
Phys.} \textbf{48} (2007) 123517 [\texttt{arXiv:0711.1103v1 [math-ph]}].

\bibitem{jayme} J. Vaz, Jr., {}{Construction of Monopoles and Instantons by
using Spinors and the Inversion Theorem}, in \emph{Clifford Algebras and
their Applications in Mathematical Physics}, V. Dietrich et al. (eds.), pp.
401-421, Kluwer, Dordrecht 1998; \emph{Clifford Algebras and Witten's
monopole equations}, in Apanasov B N, Bradlow S B, Rodrigues, Jr W A e
Uhlenbeck K K (eds.), \emph{Geometry, Topology and Physics, }de Gruyter W,
Berlin 1997.

\bibitem{jayme1} W. A. Rodrigues, Jr., Q. A. G. Souza, J. Vaz, Jr., Jr., and
P. Lounesto, \emph{Dirac-Hestenes Spinor Fields On Riemann-Cartan Manifolds}%
, Int. J. Theor. Phys. \textbf{35} (1996) 1849-1900.

\bibitem{bala} A. P. Balachandran, G. Marmo, B.-S. Skagerstam, and A. Stern, 
\emph{Magnetic monopoles with no strings}, Nucl. Phys. \textbf{B 162} (1980)
385-392.

\bibitem{balawal} W. A. Rodrigues, Jr., \emph{\ The relation between
Maxwell, Dirac and the Seiberg-Witten Equations}, Int. J. Math. and
Mathematical Sciences \textbf{2003} (2003) 2707-2734 [\texttt{%
arXiv:math-ph/0212034}].

\bibitem{benn} I. M. Benn and R. W. Tucker, \emph{\ An Introduction to
Spinors and Geometry with Applications in Physics}, Adam Hilger, Bristol
1987.

\bibitem{moro} R. A. Mosna and W. A. Rodrigues, Jr., \emph{The bundles of
algebraic and Dirac-Hestenes spinor fields}, J. Math. Phys. \textbf{45}
(2004) 2945-2988 [\texttt{arXiv:math-ph/0212033v5}]

\bibitem{rocha1} R. da Rocha and J. Vaz, Jr., \emph{Revisiting Clifford
algebras and spinors II: Weyl spinors in Cl(3,0) and Cl(0,3) and the Dirac
equation}, [\texttt{arXiv:math-ph/0412075v1}].

\bibitem{wal} W. A. Rodrigues, Jr. and E. Capelas de Oliveira, \emph{The
Many Faces of Maxwell, Dirac and Einstein Equations. A Clifford Bundle
Approach}, Lecture Notes in Physics \textbf{722}, Springer, New York 2007.

\bibitem{lawmi} {H. B. Lawson, Jr. and M. L. Michelson, \textit{Spin Geometry%
}, Princeton University Press, Princeton 1989.}

\bibitem{choquet} {Y. Choquet-Bruhat, C. DeWitt-Morette, and M.
Dillard-Bleick, \emph{Analysis, Manifolds and Physics (revised edition)},
North-Holland Publ. Co, Amsterdam 1977.}

\bibitem{rod} W. A. Rodrigues, Jr., \emph{Algebraic and Dirac Hestenes
Spinors and Spinor Fields}, {J. Math. Phys}. \textbf{45} (2004) 2908-2966 [%
\texttt{arXiv:math-ph/0212030v6.}]

\bibitem{doran} C. Doran, \emph{Geometric Algebra and its Applications to
Mathematical Physics}, Thesis, Univ. Cambridge, 1994.

\bibitem{lase} S. F. Gull, A. N. Lasenby and C. J. L. Doran, \emph{Imaginary
Numbers are not Real - the Geometric Algebra of Spacetime}, Found. Phys. 
\textbf{23} (1993) 1175-1201.

\bibitem{dor2} C. Doran, A. Lasenby A, and S. Gull, \emph{States and
operators in the spacetime algebra}, Found. Phys. \textbf{23} (1993)
1239-1264.

\bibitem{cra} J. P. Crawford, \emph{On the Algebra of Dirac Bispinor
Densities: Factorization and Inversion Theorems}, J. Math. Phys. \textbf{26}
(1985) 1429-1441; \textit{The geometric structure of the space of fermionic
physical observables}, em Micali A et al.(eds.) \textit{Clifford Algebras
and Their Applications in Math. Physics}, Kluwer Acad. Publishers, Dordrecht
1989.

\bibitem{holl} P. R. Holland, \emph{Relativistic Algebraic Spinors and
Quantum Motions in Phase Space}, {Found. Phys.} \textbf{16} (1986) 708-709.

\bibitem{hol} P. R. Holland, \emph{\ Minimal Ideals and Clifford Algebras in
the Phase Space Representation of spin-1/2 Fields}, p. 273-283 in Chisholm J
S R and Common A K (eds.), \emph{Proceedings of the Workshop on Clifford
Algebras and their Applications in Mathematical Physics (Canterbury 1985)},
Reidel, Dordrecht 1986.

\bibitem{koso} M. R. Francis and A. Kosowsky, \emph{The construction of
spinors in geometric algebra}, Annals Phys. \textbf{317} (2005) 383-409 [%
\texttt{arXiv:math-ph/0403040v2}].

\bibitem{plaga} R. Plaga, \emph{The non-equivalence of Weyl and Majorana
neutrinos with standard-model gauge interactions}, [\texttt{%
arXiv:hep-ph/0108052v1}].

\bibitem{bud} P. Budinich, \emph{From the geometry of pure spinors with
their division algebras to fermion physics}, Found. Phys. \textbf{32} (2002)
1347-1398.

\bibitem{cru} A. Crumeyrolle, \emph{Orthogonal and Symplectic Clifford
Algebras: Spinor Structures, Kluwer Academic}, Dordrecht 1990.

\bibitem{chev} C. Chevalley, \emph{The Algebraic Theory of Spinors},
Columbia University Press, New York 1954.

\bibitem{meudirac} R. da Rocha and J. Vaz, Jr., \emph{Conformal structures
and twistors in the paravector model of spacetime}, Int. J. Geom. Meth. Mod.
Phys. \textbf{4} (2007) 547-576 [\texttt{arXiv:math-ph/0412074v2}].

\bibitem{op} V. Figueiredo, E. Capelas de Oliveira, W. A. Rodrigues, Jr., 
\emph{Covariant, algebraic, and operator spinors}, Int. J. Theor. Phys. 
\textbf{29} (1990) 371-395.

\bibitem{hes} D. Hestenes, \emph{Real Spinor Fields}, J. Math. Phys. \textbf{%
8} (1967) 798-808; \textit{Spacetime Algebra}, Gordon and Breach, N. York
1966; \textit{New Foundations for Classical Mechanics}, Kluwer Acad.
Publishers, Dordrecht 1990; \emph{Observables, operators and complex numbers
in the Dirac theory}, J. Math. Phys. \textbf{16}, (1975) 556-572; \emph{%
Proper Particle Mechanics}, J. Math. Phys. \textbf{15} (1974) 1768; \emph{%
Real Spinor Fields}, J. Math. Phys. \textbf{8} (1967) 798.

\bibitem{hes1} D. Hestenes and G. Sobczyk, \emph{Clifford Algebra to
Geometric Calculus: A Unified Language for Mathematics and Physics}, D.
Reidel, Dordrecht 1984.

\bibitem{cartan} E. Cartan, \emph{The Theory of Spinors}, translated from 
\emph{Le\c{c}ons sur la th\'{e}orie des spineurs}, 1937, Dover, New York
1966.

\bibitem{riesz} M. Riesz, \emph{Clifford Numbers and Spinors}, University of
Maryland Press, College Park 1958.

\bibitem{qtmosna} R. A. Mosna and J. Vaz Jr., \emph{Quantum Tomography for
Dirac Spinors}, {Phys. Lett.} \textbf{A 315} (2003) 418-425 [\texttt{%
arXiv:quant-ph/0303072v2}].

\bibitem{jal} R. da Rocha and J. Vaz, Jr., \emph{Clifford
algebra-parametrized octonions and generalizations}, J. Algebra \textbf{301}
(2006) 459-473 [\texttt{arXiv:math-ph/0603053v1}].

\bibitem{gau} D. V. Ahluwalia, Cheng-Yang Lee, D. Schritt, T. F. Watson, {}{%
Dark matter and dark gauge fields}, in ``Dark matter in astroparticle and
particle physics, DARK 2007, Proceedings of the 6th international Heidelberg
conference'' (24-28 September 2007, Sydney, Australia), Eds. H. V.
Klapdor-Kleingrothaus and G. F. Lewis, pp. 198-208. [\texttt{%
arXiv:0712.4190v2 [hep-ph]}]; {}{Local fermionic dark matter with mass
dimension one}, [\texttt{arXiv:0804.1854v3 [hep-th]}].




\end{thebibliography}
\end{document}